\documentclass[
prl,preprint,
superscriptaddress,
showpacs,
amsmath,amssymb,
aps
]{revtex4-1}

\setcitestyle{super}

\usepackage{graphicx}
\usepackage{amsmath}
\usepackage[version=3]{mhchem}
\usepackage[usenames]{color}
\usepackage{braket}
\usepackage{color}
\usepackage{soul}
\usepackage{braket}
\usepackage{textcomp}
\usepackage{physics}
\usepackage[colorlinks=true, linkcolor=blue, citecolor=blue, urlcolor=blue,pdftex]{hyperref}
\definecolor{newcolor}{rgb}{0.9,0,0.1}

\usepackage{orcidlink}

\usepackage{bm}
\usepackage{natmove}

\newcommand{\figref}[1]{Fig.~\ref{#1}}
\newcommand{\df}[0]{{\it $\Delta\text{f}$} }

\begin{document}

\title{Charge State-Dependent Symmetry Breaking of Atomic Defects in Transition Metal Dichalcogenides}


\author{Feifei Xiang\,\orcidlink{0000-0002-4930-5919} \textsuperscript{\textdagger}}

\affiliation{nanotech@surfaces Laboratory, Empa -- Swiss Federal Laboratories for Materials Science and Technology, D\"ubendorf 8600, Switzerland}

\author{Lysander Huberich\textsuperscript{\textdagger}}
\affiliation{nanotech@surfaces Laboratory, Empa -- Swiss Federal Laboratories for Materials Science and Technology, D\"ubendorf 8600, Switzerland}

\author{Preston A. Vargas\textsuperscript{\textdagger}}
\affiliation{Department of Materials Science and Engineering, University of Florida, Gainesville, FL, 32611, USA}

\author{Riccardo Torsi}
\affiliation{Department of Materials Science and Engineering, The Pennsylvania State University, University Park, PA 16082, USA}

\author{Jonas Allerbeck\,\orcidlink{0000-0002-3912-3265}}
\affiliation{nanotech@surfaces Laboratory, Empa -- Swiss Federal Laboratories for Materials Science and Technology, D\"ubendorf 8600, Switzerland}

\author{Anne Marie Z. Tan}
\affiliation{Department of Materials Science and Engineering, University of Florida, Gainesville, FL, 32611, USA}
\affiliation{Institute of High Performance Computing (IHPC), Agency for Science, Technology and Research (A*STAR), Singapore 138632, Republic of Singapore}

\author{Chengye Dong}
\affiliation{Two-Dimensional Crystal Consortium, The Pennsylvania State University, University Park, PA 16802, USA}

\author{Pascal Ruffieux\,\orcidlink{0000-0001-5729-5354}}
\affiliation{nanotech@surfaces Laboratory, Empa -- Swiss Federal Laboratories for Materials Science and Technology, D\"ubendorf 8600, Switzerland}

\author{Roman Fasel\,\orcidlink{0000-0002-1553-6487}}
\affiliation{nanotech@surfaces Laboratory, Empa -- Swiss Federal Laboratories for Materials Science and Technology, D\"ubendorf 8600, Switzerland}

\author{Oliver Gr\"oning}
\affiliation{nanotech@surfaces Laboratory, Empa -- Swiss Federal Laboratories for Materials Science and Technology, D\"ubendorf 8600, Switzerland}

\author{Yu-Chuan Lin}
\affiliation{Department of Materials Science and Engineering, The Pennsylvania State University, University Park, PA 16082, USA}
\affiliation{Department of Materials Science and Engineering, National Yang Ming Chiao Tung University, Hsinchu City 300, Taiwan}

\author{Richard G. Hennig\,\orcidlink{0000-0003-4933-7686}}
\affiliation{Department of Materials Science and Engineering, University of Florida, Gainesville, FL, 32611, USA}

\author{Joshua A. Robinson\,\orcidlink{0000-0002-1513-7187}}
\affiliation{Department of Materials Science and Engineering, The Pennsylvania State University, University Park, PA 16082, USA}
\affiliation{Two-Dimensional Crystal Consortium, The Pennsylvania State University, University Park, PA 16802, USA}
\affiliation{Department of Chemistry and Department of Physics, The Pennsylvania State University, University Park, PA, 16802, USA}

\author{Bruno Schuler\,\orcidlink{0000-0002-9641-0340}}
\email[]{bruno.schuler@empa.ch}
\affiliation{nanotech@surfaces Laboratory, Empa -- Swiss Federal Laboratories for Materials Science and Technology, D\"ubendorf 8600, Switzerland}

\begin{abstract} 
\textbf{
The functionality of atomic quantum emitters is intrinsically linked to their host lattice coordination. Structural distortions that spontaneously break the lattice symmetry strongly impact their optical emission properties and spin-photon interface.
Here we report on the direct imaging of charge state-dependent symmetry breaking of two prototypical atomic quantum emitters in mono- and bilayer MoS$_2$ by scanning tunneling microscopy (STM) and non-contact atomic force microscopy (nc-AFM).
By substrate chemical gating different charge states of sulfur vacancies (Vac$_\text{S}$) and substitutional rhenium dopants (Re$_\text{Mo}$) can be stabilized. Vac$_\text{S}^{-1}$ as well as Re$_\text{Mo}^{0}$ and Re$_\text{Mo}^{-1}$ exhibit local lattice distortions and symmetry-broken defect orbitals attributed to a Jahn-Teller effect (JTE) and pseudo-JTE, respectively. 
By mapping the electronic and geometric structure of single point defects, we disentangle the effects of spatial averaging, charge multistability, configurational dynamics, and external perturbations that often mask the presence of local symmetry breaking.
}
\end{abstract}

\date{\today}
\maketitle

Mastering the controlled introduction of defects and impurities in semiconductors has proven pivotal to their transformative technological success. 
As miniaturization reaches atomic length scales and quantum applications emerge, stochastic distributions make way for single impurity engineering~\cite{philips2022universal,kiczynski2022engineering}. This need for precise defect manipulation extends to the realm of artificial atom qubits, a cornerstone of quantum sensing and quantum communication~\cite{awschalom2018quantum,atature2018material,awschalom2018quantum}.
Atomic quantum emitters based on defects in solids, rely on spin-selective optical decay pathways enabling the high-fidelity spin initialization and readout even at room temperature~\cite{dolde2013room}. 
Two-dimensional (2D) materials such as hexagonal boron nitride or transition metal dichalcogenides (TMDs) emerged as a game-changing platform to host such quantum emitters because they lack surface dangling bonds that degrade spin coherence or photon distinguishability~\cite{tran2016quantum,gottscholl2020initialization}. Furthermore, 2D materials exhibit increased extraction efficiency~\cite{aharonovich2016solid}, integrate seamlessly with quantum optical devices~\cite{wu2015monolayer,liu20192d}, and support electrical fine-tuning~\cite{grosso2017tunable,noh2018stark} and spatial engineering~\cite{liu20192d,klein2019site} of defects and their associated electronic states. Regardless of the host material, the local defect symmetry governs the level structure and optical selection rules that ultimately provide the spin-photon interface. Spontaneous symmetry breaking from Jahn-Teller distortions and strain fields can therefore dramatically change the radiative recombination rates, emission wavelength and polarization~\cite{ulbricht2016jahn,lee2022spin}. Jahn-Teller systems, in particular, can strongly influence spin-orbit coupling that gives rise to the inter-system crossing, which yields the spin contrast~\cite{thiering2017ab,streltsov_jahn-teller_2020,lee2022spin}.
Therefore, understanding the mechanisms that break the coordination symmetry of single atom qubits as a function of their charge state and external factors such as strain are decisive to tailor their functionality.\\

Although it is well known that TMDs and other transition metal compounds exhibit enhanced susceptibility towards Jahn-Teller effect (JTE), detecting JTE experimentally is challenging, because configurational dynamics and spatial averaging can obscure the minute local distortions~\cite{bersuker_jahnteller_2021}. Additionally, these distortions are often affected by external perturbation, which can complicate the interpretation of experimental results~\cite{bersuker_jahnteller_2021}. Therefore, even though the JTE and its extensions~\cite{bersuker2023four} has been discovered almost a century ago, its abundance and significance in different materials systems has only gradually been revealed and is still mostly inferred theoretically by \textit{ab initio} methods~\cite{gali2019ab,bersuker_jahnteller_2021}. While a number of TMD point defects including chalcogen vacancies~\cite{komsa_native_2015,tan_stability_2020} and substitutional transition metal dopants~\cite{cheng_prediction_2013} have been predicted to undergo spontaneous symmetry breaking, it has only rarely been verified experimentally using aberration-corrected transmission electron microscopy~\cite{bertoldo_intrinsic_2021}, and no direct experimental observation of symmetry-broken electronic states has been reported to date.\\

Here we present direct experimental and theoretical evidence for a Jahn-Teller driven symmetry breaking of chalcogen vacancies~\cite{klein_site-selectively_2019} and rhenium-based quantum emitters~\cite{gupta_two-level_2019} in MoS$_2$.
Scanning tunneling microscopy (STM) orbital imaging~\cite{repp_molecules_2005}, CO-tip noncontact atomic force microscopy (nc-AFM) measurements~\cite{gross_chemical_2009}, and density functional theory (DFT) reveal the symmetry-broken defect orbitals and the structural distortion for negatively charged sulfur vacancy (Vac$_\text{S}^{-1}$) and rhenium substituting for molybdenum in its neutral and negative charge state (Re$_\text{Mo}^0$, Re$_\text{Mo}^{-1}$).
We reveal that Vac$_\text{S}^{-1}$ coexist in both a symmetric and distorted geometry, which may explain why the most studied TMD defect has evaded the spectroscopic detection of spontaneous symmetry breaking so far.
Charge state tristability of Re dopants is achieved by chemical gating \textit{via} the underlying substrate. 
We find larger domains where the structural distortions are aligned, which may indicate a non-negligible strain field in the epitaxially-grown MoS$_2$ samples. 
Surprisingly, the comparatively broad defect resonances of some Re impurities exhibit a continuous transformation of the measured defect orbital, suggesting a configurational continuum as a consequence of the flat potential energy surface experienced by the Re impurity.

\section*{Jahn-Teller Driven Symmetry-Breaking of Vac$_\text{S}^{-1}$ in MoS$_2$}
Chalcogen vacancies are the most prevalently discussed point defect in TMDs due to their low formation energy~\cite{komsa_native_2015}, frequent detection in transmission electron microscopy~\cite{komsa_two-dimensional_2012,zhou_intrinsic_2013}, and attributed sub-band gap emission in optical spectroscopy~\cite{tongay_defects_2013,mitterreiter_role_2021}. Despite their disputed presence at ambient conditions due to their high reactivity~\cite{barja_identifying_2019}, recent strategies involving deliberate vacancy generation by annealing or ion bombardment and protection by vacuum or inert capping layers made it possible to directly measure unpassivated vacancies in TMDs~\cite{schuler_large_2019,klein_site-selectively_2019,mitterreiter_role_2021}.\\

\begin{figure*}[h]
\includegraphics[width=\textwidth]{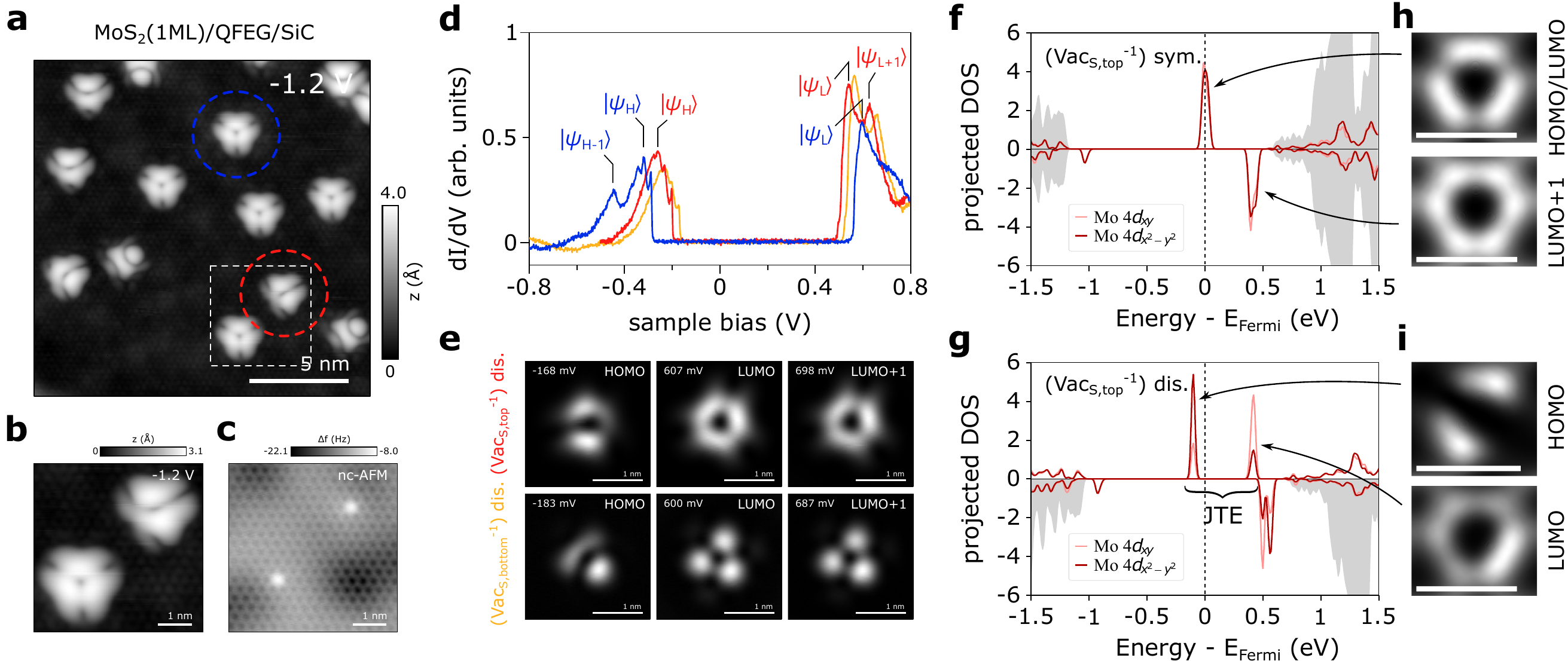}
\caption{\label{fig:Vac_S}
\textbf{Symmetric and symmetry-broken (distorted) negatively charged sulfur vacancies (Vac$_\text{S}^{-1}$).}
\textbf{a}, STM topography of annealing-induced Vac$_\text{S}^{-1}$ (colored circles) in monolayer MoS$_2$/QFEG/SiC. 
\textbf{b}, Zoom-in STM topography of the white dashed box in (a) containing a distorted (top right) and a symmetric (bottom left) Vac$_\text{S}^{-1}$.
\textbf{c}, Nc-AFM image with a metallic tip of the same area as in (b) proving that both defects are Vac$_\text{S}$ in the upper S layer.
\textbf{d}, d$I$/d$V$ spectra of the frontier in-gap defect orbitals of the symmetric (blue) and distorted (red) Vac$_\text{S, top}^{-1}$, as well as the distorted Vac$_\text{S, bottom}^{-1}$ (orange) ($V_\text{mod}$ = 2\,mV). 
\textbf{e}, Corresponding constant height dI/dV maps of the distorted S top (upper row) and bottom (bottom row) vacancies highest occupied (HOMO) and lowest unoccupied defect orbitals (LUMO, LUMO+1), labelled in d ($V_\text{mod}$ = 20\,mV). 
\textbf{f,g}, Projected density of states (pDOS) of Vac$_\text{S}^{-1}$ in MoS$_2$ in the symmetrized (f) and distorted (g, ground state) geometry, calculated in a 4$\times$4 supercell. The pDOS onto the Mo $d_{xy}$ and $d_{x^2-y^2}$ orbitals are shown in red. 
\textbf{h,i}, Constant height DOS map of the HOMO and LUMO(+1) orbitals for the symmetrized (h) and diststorted (i) Vac$_\text{S, top}^{-1}$ geometry
}
\end{figure*}

\figref{fig:Vac_S}a shows a STM image of monolayer MoS$_2$ grown by metal-organic chemical vapor deposition (MOCVD) on quasi-freestanding epitaxial graphene on 6H-silicon carbide (0001) (QFEG). S vacancies (Vac$_\text{S}$) were induced by annealing the sample up to 900\,°C in ultrahigh-vacuum~\cite{schuler_large_2019}. Apart from common (unintentional) as-grown defects like O$_\text{S}$, CH$_\text{S}$, Cr$_\text{Mo}$, and W$_\text{Mo}$~\cite{schuler_how_2019} (see Fig.~S2) that are also present at much lower annealing temperatures, we find the anticipated Vac$_\text{S}$ in both the upper and lower S layer as previously identified~\cite{schuler_large_2019,mitterreiter_atomistic_2020}. All S vacancies are negatively charged because the QFEG substrate sets the Fermi level relatively high up in the MoS$_2$ band gap, such that the lowest Vac$_\text{S}$ in-gap state becomes occupied. The negative charge can be inferred from the band bending-induced dark halo around the defect at positive sample bias~\cite{aghajanian_resonant_2020}.\\

While some Vac$_\text{S}^{-1}$ appear threefold symmetric, others unexpectedly exhibit a `distorted' appearance with only a two-fold symmetry at negative sample bias (\figref{fig:Vac_S}a,b), even though both type of defects appear in nc-AFM as sulfur vacancies in the top sulfur layer (\figref{fig:Vac_S}c).  Also in scanning tunneling spectroscopy (STS) measurements the electronic structure between the symmetric and distorted vacancy species are distinct as shown in \figref{fig:Vac_S}d. The symmetric Vac$_\text{S}^{-1}$ has two occupied in-gap defect states (HOMO-1 and HOMO) and one LUMO resonance close to the Fermi energy (blue), while the distorted Vac$_\text{S}^{-1}$ features one HOMO resonance (with vibronic side-bands) and two LUMO resonances (red and orange). The symmetric Vac$_\text{S}^{-1}$ preserves the underlying C$_{3v}$ lattice symmetry in its frontier orbitals (see Fig.~S4e,f), whereas the HOMO of the distorted Vac$_\text{S}^{-1}$ variant has a lower C$_{2v}$ symmetry (see \figref{fig:Vac_S}e). Such a behavior was not observed for Vac$_\text{S}^0$ in WS$_2$~\cite{schuler_large_2019}, which indicates the decisive role of the defect charge state.
\\

We attribute the reduced point group symmetry of Vac$_\text{S}^{-1}$ to a JTE as previously predicted~\cite{tan_stability_2020}. Upon charging, the degenerated S vacancy state (\figref{fig:Vac_S}f) becomes unstable and relaxes, thereby lifting the orbital degeneracy (\figref{fig:Vac_S}g). The distorted Vac$_\text{S}^{-1}$ geometry (Fig.~S5c-e) has a lower total energy by 116\,meV~\cite{tan_stability_2020}. The calculated density of states (DOS) map of the frontier orbitals shown in \figref{fig:Vac_S}i are in excellent agreement with the experimental differential conductance (d$I$/d$V$) maps.
However, the reason for the co-existence of both the symmetric and distorted Vac$_\text{S}^{-}1$ is not entirely clear. While a doubly negatively charged Vac$_\text{S}^{-2}$ would have a symmetric ground state~\cite{tan_stability_2020}, it is unlikely to be formed due to the large Coulomb energy associated with the highly confined defect wave function, in-line with the calculated charge transition level from -1 to -2, which is above the conduction band minimum~\cite{tan_stability_2020}.
Possibly, the system is trapped in a local minimum with a potential barrier too high to relax to the global potential energy minimum.\\

Motivated by the surprising discovery of symmetry breaking in one of the best studied defects in TMDs, we designed a defect system that we can stabilize in different charge states by chemical gating \textit{via} the substrate. Next we will discuss the charge state tristability of rhenium dopants in MoS$_2$ and their charge state-dependent symmetry breaking. Rhenium substituted for molybdenum (Re$_\text{Mo}$) has a D$_{3h}$ point symmetry group and produces fundamentally different in-gap states as compared to Vac$_\text{S}$, which makes it an ideal candidate to test the generality of our findings.

\section*{Charge State Tristability of Re Dopants in MoS$_2$}
Chemical doping of 2D semiconductors has been a popular but challenging route to control their conductivity due to the large defect ionization energies associated with the tightly confined defect wave functions in 2D~\cite{zhu_dimensionality-inhibited_2021}. Recently, a scalable MOCVD process has been developed to introduce Re dopants in TMDs in controllable concentrations from hundreds of ppm to percentage concentrations~\cite{kozhakhmetov_scalable_2020,torsi_dilute_2023}. Substitutional Re dopants in WSe$_2$ on QFEG are positively charged (ionized) and feature a series of closely-spaced unoccupied defect states below the conduction band minimum.
Here we leverage the higher electron affinity $\chi$ of monolayer MoS$_2$ as compared to WSe$_2$, to position the Fermi level in-between these Re$_\text{Mo}$ defect states. In MoS$_2$/QFEG, we find that most Re impurities are charge neutral while some remain positively charged, which also applies to Re-doped bilayer MoS$_2$ grown on QFEG.
By using epitaxial graphene (EG) without hydrogen intercalation as a substrate, which has a substantially smaller work function than QFEG, we push the Fermi level even higher\cite{pan_wse2_2019,subramanian_tuning_2020}. Counter-intuitive for an $n$-type dopant, this results in the Re impurities becoming negatively charged, i.e. they accept an additional electron. 
By substrate chemical gating, we establish a charge state tristability of Re dopants in MoS$_2$.\\

\begin{figure*}[h]
\includegraphics[width=\textwidth]{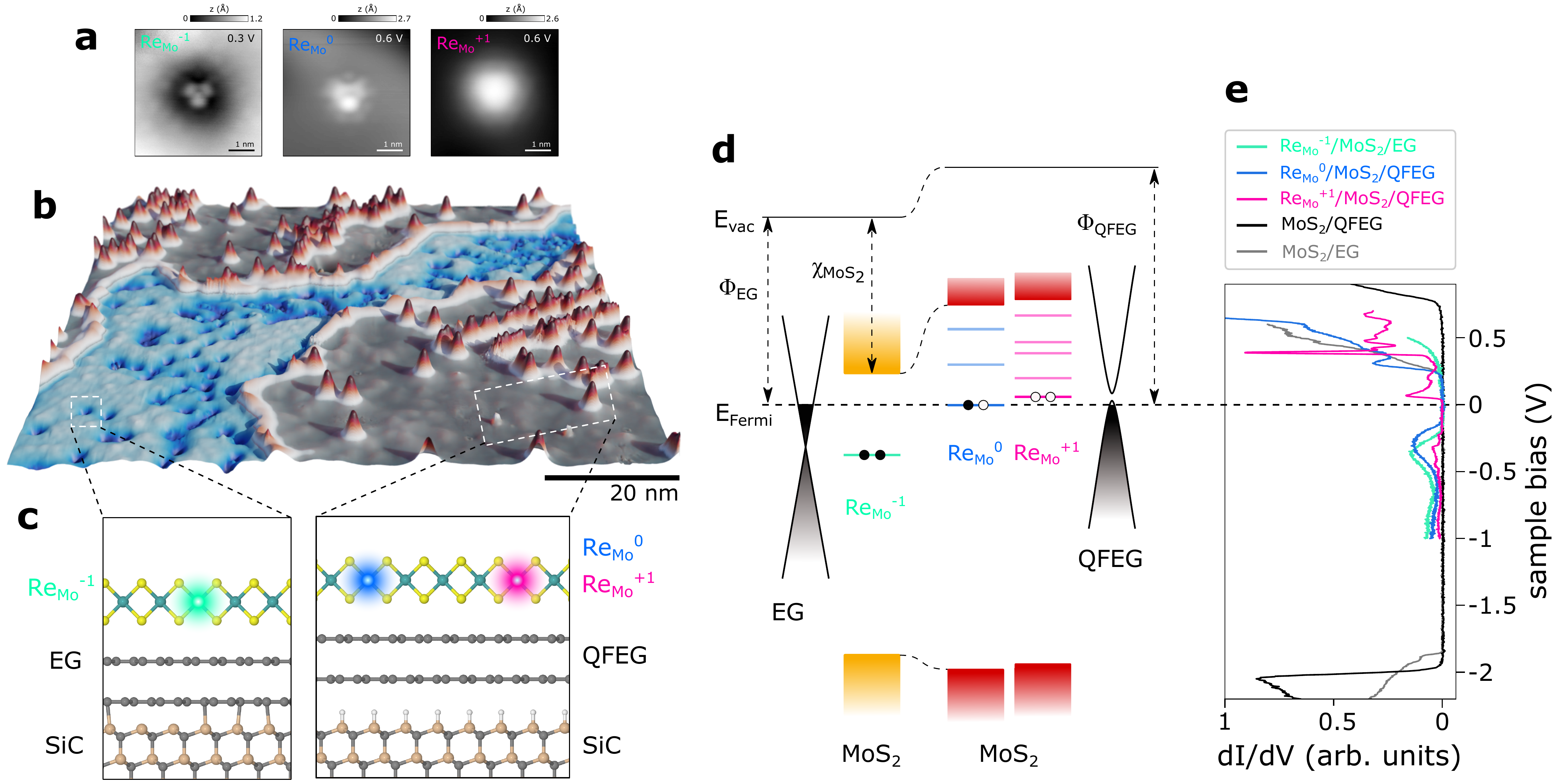}
\caption{\label{fig:charge_tristability}
\textbf{Charge state tristability of Re dopants in MoS$_2$ on EG and QFEG on SiC.}
\textbf{a}, STM topographies of single Re dopants in the three different charge states (Re$_\text{Mo}^{-1}$, Re$_\text{Mo}^{0}$, and Re$_\text{Mo}^{+1}$).
\textbf{b} 3D representation of a large-scale STM image of Re-doped monolayer MoS$_2$ on domains of EG (center blue region) or domains of QFEG substrate (outer gray region). 
At positive bias (here 0.6\,V), Re dopants are imaged as a dark blue pits on EG whereas they appear as red hills on MoS$_2$ grown on QFEG, depending on their charge states.
\textbf{c}, Atomic model of the MoS$_2$/EG and MoS$_2$/QFEG interface. On EG only Re$_\text{Mo}^{-1}$ are observed, whereas on QFEG both Re$_\text{Mo}^{0}$ and Re$_\text{Mo}^{+1}$ are found.
\textbf{d}, Schematic electronic level diagram of the MoS$_2$/EG (left) and MoS$_2$/QFEG interface (right) with the Re$_\text{Mo}$ level occupation indicated. The work function of EG $\Phi_\text{EG}$ is substantially smaller than the work function of QFEG $\Phi_\text{QFEG}$. $\chi_\text{MoS$_2$}$ denotes the electron affinity of MoS$_2$.
\textbf{e}, dI/dV spectra of Re$_\text{Mo}^{-1}$ (green), Re$_\text{Mo}^{0}$ (blue), and Re$_\text{Mo}^{+1}$ (pink), pristine monolayer MoS$_2$ on EG (grey) and on QFEG (black), respectively. Bias modulation amplitude: $V_\text{mod}$ = 20\,mV.
}
\end{figure*}

In \figref{fig:charge_tristability}b we present a large-scale STM topography of Re-doped (5\%) MoS$_2$ on partially H-intercalated EG. In areas with H-intercalation (gray plateau in \figref{fig:charge_tristability}b and structural model shown in \figref{fig:charge_tristability}c), we observe the Re dopants in the positive (Re$_\text{Mo}^{+1}$) or neutral (Re$_\text{Mo}^{0}$) charge state, whereas on EG substrates (blue canyon in \figref{fig:charge_tristability}b, and structural model in \figref{fig:charge_tristability}c) all Re are negatively charged (Re$_\text{Mo}^{-1}$). The charge states can be discriminated by the characteristic upwards (anionic impurity) or downwards (cationic impurity) band bending in STS. This translates into a dark depression (\figref{fig:charge_tristability}a left) or bright protrusion (\figref{fig:charge_tristability}a right) of the STM topography if measured at the conduction band edge.
Most importantly, Re$_\text{Mo}^{+1}$ preserves the underlying D$_{3h}$ lattice symmetry, as previously reported for Re$_\text{W}^{+1}$ in WSe$_2$~\cite{kozhakhmetov_scalable_2020}, while the STM topography of Re$_\text{Mo}^{0}$ and Re$_\text{Mo}^{-1}$ appear slightly distorted.\\

The pink d$I$/d$V$ spectrum in \figref{fig:charge_tristability}e and defect orbitals depicted in Fig.~S6 for Re$_\text{Mo}^{+1}$ closely resemble the Re$_\text{W}^{+1}$ states observed in WSe$_2$~\cite{kozhakhmetov_scalable_2020}, with several defect states just above the Fermi level. Re$_\text{Mo}^{0}$ (\figref{fig:charge_tristability}e blue, and Fig.~S7) exhibits several unoccupied defect states above Fermi, along with an unusually broad defect resonance below it. Lastly, Re$_\text{Mo}^{-1}$ shown in green in \figref{fig:charge_tristability}e and Fig.~S8, has only a broad defect resonance below the Fermi level.
Based on the STS spectra we derive a simplified schematic level diagram for the differently charged Re$_\text{Mo}$ impurities in \figref{fig:charge_tristability}d.
The EG and QFEG act as a Fermi sea where electrons can be donated to or withdrawn from the Re$_\text{Mo}$. Upon (de)charging, many-body effects renormalize the energies of all defect states because of their pronounced localization. For instance, the unoccupied defect states for Re$_\text{Mo}^{-1}$ are pushed higher in energy, whereas they become closely spaced at low energy for Re$_\text{Mo}^{+1}$.
\\

\section*{Symmetry Breaking of Re$_\text{Mo}^{-1}$ and Re$_\text{Mo}^0$}
\begin{figure*}[h]
\includegraphics[width=\textwidth]{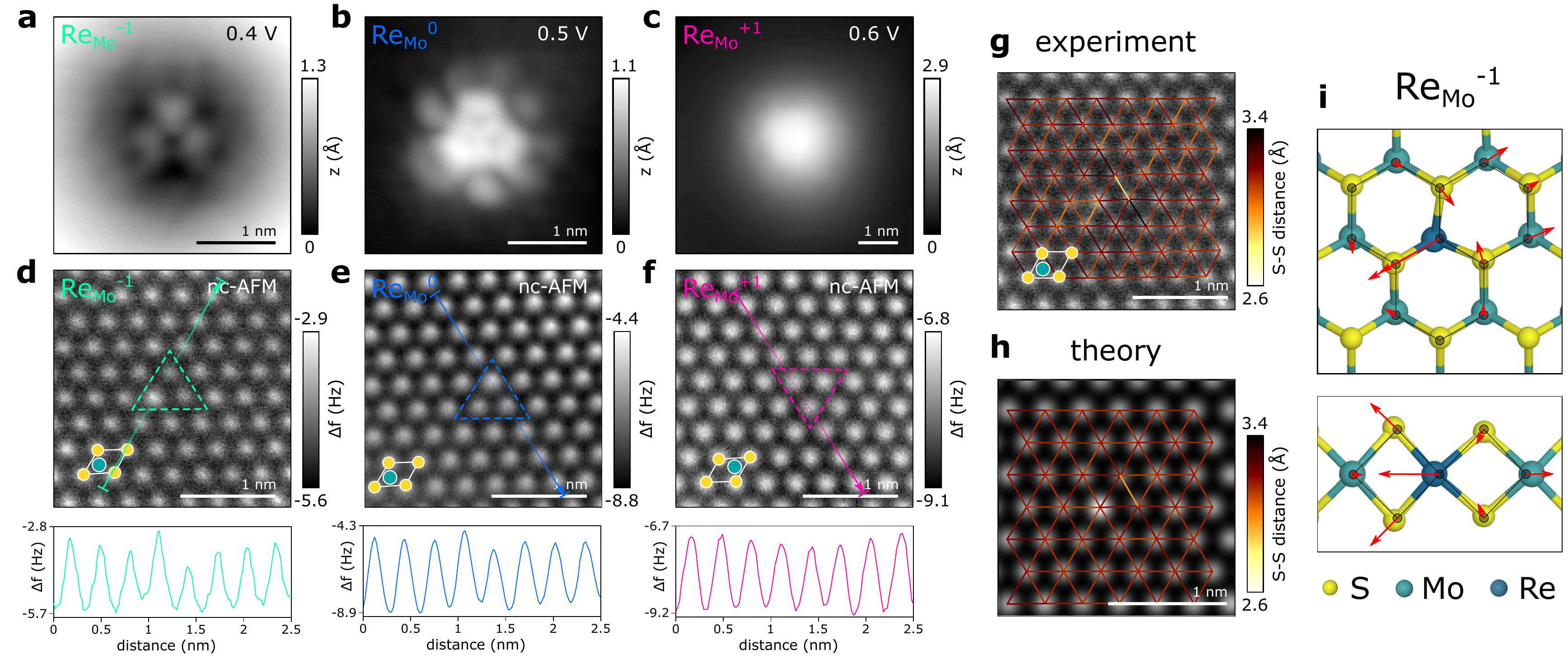}
\caption{\label{fig:AFM_strain}
\textbf{Charge state-dependent Jahn-Teller distortions.}
\textbf{a-c}, STM topography of Re$_\text{Mo}^{-1}$, Re$_\text{Mo}^{0}$, and Re$_\text{Mo}^{+1}$, respectively. \textbf{d-f}, CO-tip nc-AFM image of the same Re impurity as shown in a-c. Top sulfur atoms are repulsive (bright) and molybdenum atoms attractive (dark). The MoS$_2$ unit cell is displayed at the bottom left in each image (S: yellow, Mo: turquoise). The Re atoms are located in the center of the dashed triangle, bound to the three bright S atoms on the top surface. A \df line profile across a sulfur row (arrow in image) is shown in the lower panel, highlighting a protruding S atom next to the Re impurity for Re$_\text{Mo}^{-1}$ and slightly weaker for Re$_\text{Mo}^{0}$. 
\textbf{g,h}, Measured (g) and simulated (h) CO-tip nc-AFM image of Re$_\text{Mo}^{-1}$. The apparent sulfur-sulfur distances in the top S layer is indicated by the red-to-yellow lines. Measurement and simulation show significant lateral compressive strain introduced by the Re impurity at nearest-neighbor S atoms opposite to the protruding S atom.
\textbf{i}, Calculated geometry of Re$_\text{Mo}^{-1}$ exhibiting a considerable distortion. The red arrows indicate the atom relaxations as compared to the pristine MoS$_2$ lattice (gray ball and stick model). For clarity, the arrows are ten times longer than the actual relaxation.}
\end{figure*}

To probe whether the striking electronic asymmetry of Re$_\text{Mo}^{0}$ and Re$_\text{Mo}^{-1}$ results from a (measurable) lattice distortion, we use CO-tip nc-AFM~\cite{gross_chemical_2009} as shown in \figref{fig:AFM_strain}d-f. Due to the high surface sensitivity of CO-tip nc-AFM, only the surface S atoms can be resolved as bright (repulsive) features. The Mo atom positions can be inferred from the slightly more attractive interaction as compared to the hollow site, establishing the full unit cell of 1H-MoS$_2$ indicated as an inset in \figref{fig:AFM_strain}d-f. 
Although the Re impurity itself cannot be directly resolved, its location (marked by the colored triangle) can be deduced from correlation with the STM topography (\figref{fig:AFM_strain}a-c). For Re$_\text{Mo}^{0}$ and Re$_\text{Mo}^{-1}$ we find that one of the neighboring S atoms appears brighter, indicating an outwards relaxation from the S plane, which can be seen in the line-cut below the respective nc-AFM image. Occasionally, the protruding S atom is found to switch during nc-AFM measurements at close tip--sample distance (Fig.~S9). 
For Re$_\text{Mo}^{+1}$ no such relaxation could be resolved. Apart from the vertical relaxation, which is most pronounced for Re$_\text{Mo}^{-1}$, also a lateral relaxation of the two opposing S atoms is observed (\figref{fig:AFM_strain}g). The vertical protrusion and lateral strain profile are in excellent agreement with the simulated CO-tip nc-AFM image (\figref{fig:AFM_strain}h) based on the calculated relaxed geometry shown in \figref{fig:AFM_strain}i.~\cite{hapala_mechanism_2014}\\

\begin{figure*}[h]
\includegraphics[width=\textwidth]{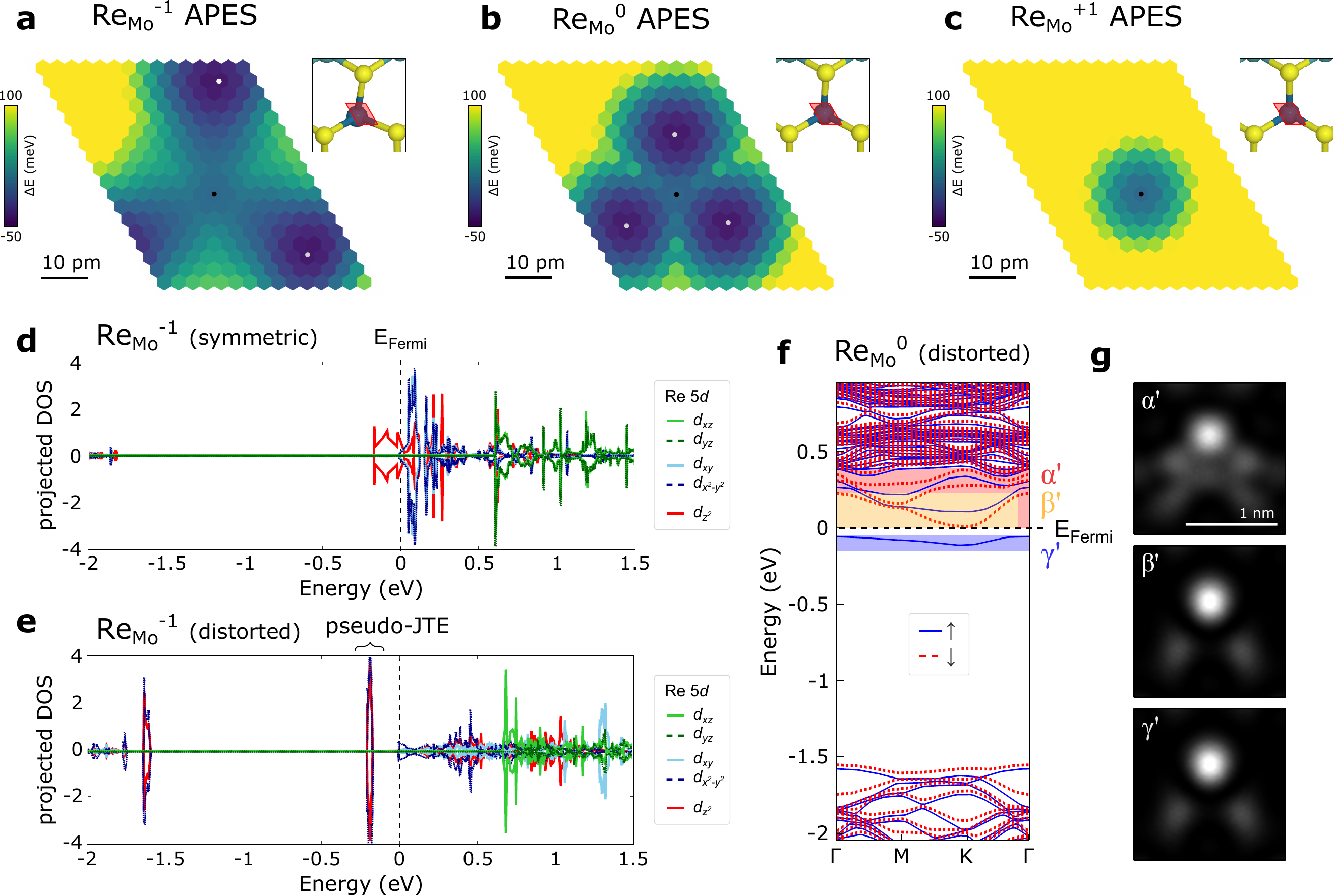}
\caption{\label{fig:theory}
\textbf{Adiabatic potential energy surface (APES) and projected DOS (pDOS) of Re$_\text{Mo}$ in different charge states.}
\textbf{a-c}, Potential energy surface of the Re dopant for the negative, neutral, and positive charge state respectively. The energies correspond to the $\Gamma$ point energy in a $5\times5$ supercell. The black and white dots indicate the symmetric and distorted positions of Re, respectively. The sampled area for the APES is marked by a red parallelogram in each structural model inset.
\textbf{d,e}, Projected density of states of Re$_\text{Mo}^{-1}$ in the constrained symmetric (d) and distorted ground state geometry (e), exhibiting electronic reconfiguration resulting from the pseudo-JTE.
\textbf{f}, Spin-polarized Re$_\text{Mo}^{0}$ band structure resulting in a paramagnetic ground state.
\textbf{g}, Constant height DOS maps of different energy windows in the Re$_\text{Mo}^{0}$ band structure, as indicated in f. The $\alpha'$ image reflects the STM contrast at low bias. The $\beta'$ and $\gamma'$ image corresponds to the lowest unoccupied and highest occupied Re$_\text{Mo}^{0}$ orbital, respectively (cf. \figref{fig:STS_continuum_directional}b C3 and D1). 
}
\end{figure*}

We employ DFT with a 2D charge correction scheme~\cite{freysoldt2018first} to investigate the atomic and electronic structure of Re$_{\text{Mo}}$ in MoS$_2$, considering the positive, neutral, and negative charge states.  While earlier \textit{ab initio} studies failed to capture the symmetry breaking in the neutral defect state~\cite{Lin2014,Zhao2017,Hallam2017}, our analysis demonstrates that enforcing integer electronic occupation successfully reproduces a distorted ground state geometry for Re$_\text{Mo}^{0}$ and Re$_\text{Mo}^{-1}$. In \figref{fig:theory}a-c, we present the adiabatic potential energy surface (APES) for all three charge states. In the positive charge state, the APES exhibits a single minimum at the high symmetry position. However, for Re$_\text{Mo}^{0}$ and Re$_\text{Mo}^{-1}$, three equivalent off-centered minima are observed at distances of 13\,pm and 25\,pm, respectively, towards one of the neighboring S atoms. Although the structural relaxation of Re away from the center is twice as large for the negative charge state, the energy gain of approximately -38\,meV is comparable for both neutral and negative charge states. 
The obtained \textit{ab initio} APES supports the experimentally observed symmetry-broken ground states of Re$_\text{Mo}^{0}$ and Re$_\text{Mo}^{-1}$.\\

To gain insight into the driving force behind the observed symmetry breaking, we examine the orbital-projected density of states (pDOS) of Re$_\text{Mo}^{-1}$ in both the symmetric and distorted ground state geometries, as shown in \figref{fig:theory}d and e, respectively. In contrast to the Vac$_\text{S}^{-1}$ case, the symmetric configuration of Re$_\text{Mo}^{-1}$ does not occupy a degenerate orbital at the Fermi energy. Instead, two electrons occupy the the non-degenerate $d_{z^2}$ HOMO (red curve in \figref{fig:theory}d). The excited state spectrum reveals degeneracies with dominant contributions from $d_{xz}$ and $d_{yz}$ (green solid and dashed line) orbitals and states composed of $d_{xy}$ and $d_{x^2-y^2}$ (blue solid and dashed line) orbitals. 
The structural distortion lowers the energy of the HOMO by creating a hybridized $d_{z^2} - d_{x^2-y^2}$ orbital that is doubly occupied and shifts down to the mid-gap region. 
\textit{A priori} it is unclear if the hybridized orbital results from occupying a $d_{z^2}$ state that takes on $d_{x^2-y^2}$ orbital character, or vice versa, which would either suggest that the mechanism is attributed to a pseudo-JTE or hidden-JTE, respectively.
To differentiate between a pseudo-JTE and a hidden-JTE, one can look at the energy along the distortion direction in the APES (\figref{fig:theory}a). A hidden-JTE would typically manifest as an energy barrier between the symmetric and distorted geometry, with the initially occupied state experiencing an energy increase that is eventually surpassed by the energy gain of the previously unoccupied state~\cite{bersuker2023four}. 
Consequently, the absence of such an energy barrier confidently identifies the distortion mechanism in this system as being governed by the pseudo-JTE.\\

Similarly, the distortion in the neutral charge state is also caused by the pseudo-JTE (cf. Fig.~S13). In the neutral charge state, the odd number of electrons gives rise to a spin-polarized ground state, wherein a singly occupied $d_{z^2} - d_{x^2-y^2}$ SOMO is established (\figref{fig:theory}f). This leads to a net magnetic moment that is absent in the positive or negative charged defects. The presence of a paramagnetic ground state in the neutral defect with a singly occupied spin channel is not captured by previous \textit{ab initio} descriptions of Re$_\text{Mo}^0$ in MoS$_2$~\cite{Zhao2017}. However, experimental studies using electron paramagnetic resonance have successfully detected Re doping in natural MoS$_2$ samples, providing further validation for our findings~\cite{Stesmans2020}. Furthermore, the spatial distribution of the highest occupied ($\gamma'$) and lowest unoccupied ($\beta'$) Re$_\text{Mo}^{0}$ orbitals shown in \figref{fig:theory}g exhibit the same spatial distribution in perfect agreement with the experimental d$I$/d$V$ images, as shown in \figref{fig:STS_continuum_directional}b (resonances C3 and D1). Therefore, we conclude that the symmetry-broken paramagnetic ground state observed in the neutral Re$_\text{Mo}$ substitution reconciles our theoretical predictions with experimental observations based on scanning probe microscopy and previous measurements using electron paramagnetic resonance.\\

\section*{Configurational Continuum and Directional Alignment}
The unusually broad defect resonances on the order of 300\,meV observed for some Re$_\text{Mo}^{0}$ and Re$_\text{Mo}^{-1}$ states raises the question about the broadening mechanism. For many TMD defects, the intrinsic lifetime broadening primarily caused by the weakly interacting graphene substrate, is only a few meV~\cite{schuler_large_2019}. We do not expect a qualitatively different substrate interaction for Re dopants, as it is mainly influenced by the distance to the substrate, and the measured defect state broadening does not decrease for Re in bilayer MoS$_2$.
In low-temperature tunneling spectroscopy of single defects in 2D semiconductors, significant broadening of electronic resonances can occur due to strong electron-phonon coupling~\cite{cochrane_spin-dependent_2021}.
However, in most cases the confined electronic states couple strongly only to a small set of localized modes with energies of a few tens of meV, given the weight of the atoms involved~\cite{cochrane_spin-dependent_2021}. Therefore, it is typically possible to differentiate the vibronic ground state resonance (zero-phonon line) and the first few vibronic excited states with an intrinsic broadening of only 3-4\,meV, as can be seen for instance in the d$I$/d$V$ spectra of the sulfur vacancy in \figref{fig:Vac_S}d. Hence, vibronic broadening is unlikely the main reason for the large defect state broadening.\\

\begin{figure*}[h]
\includegraphics[width=\textwidth]{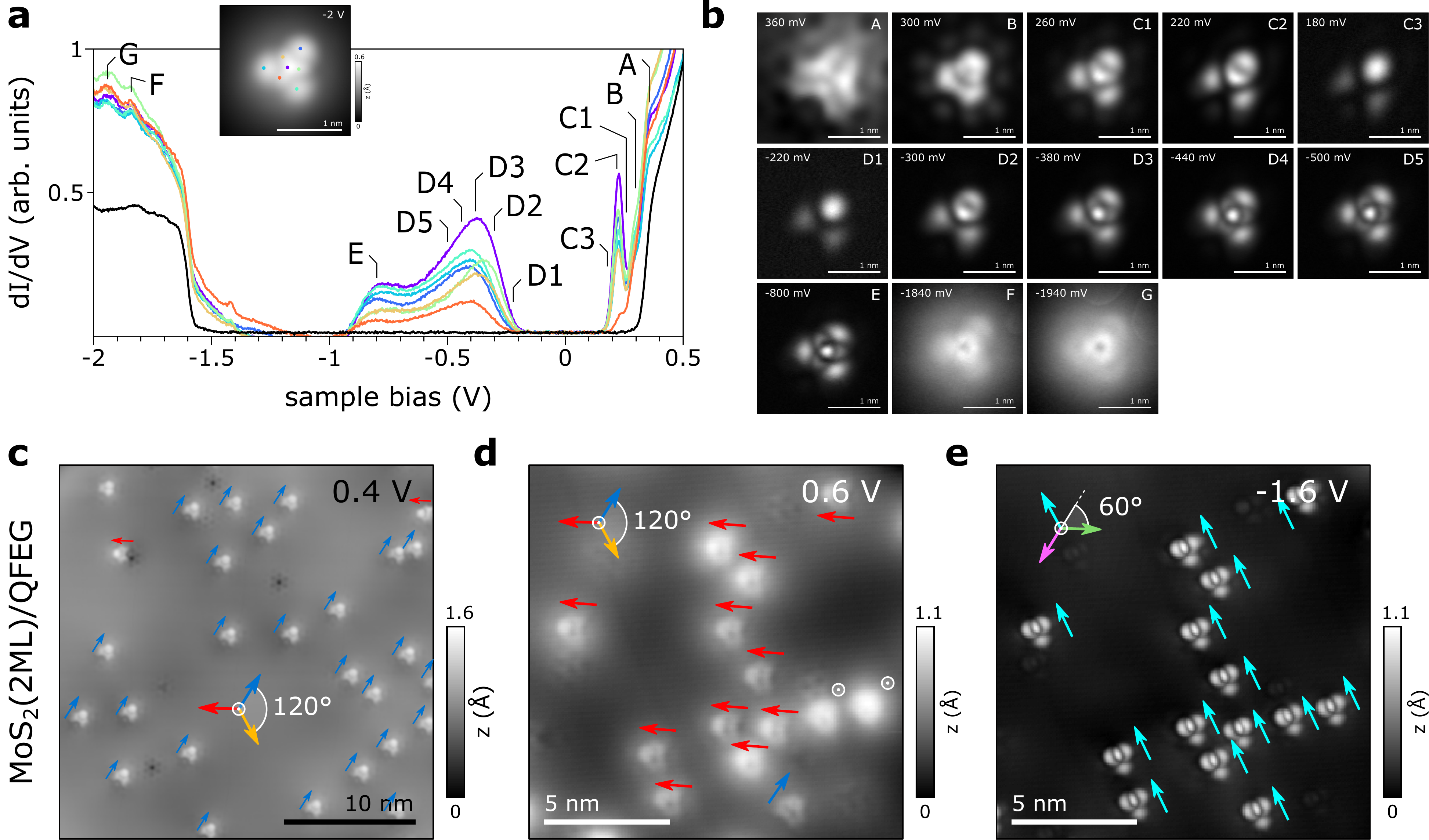}
\caption{\label{fig:STS_continuum_directional}
\textbf{Configurational continuum and directionally aligned domains.}
\textbf{a}, dI/dV spectroscopy of Re$_\text{Mo}^{0}$ on bilayer MoS$_2$ ($V_\text{mod}$ = 10\,mV). The spectroscopy locations on the defect are indicated with a dot of the same color in the inset STM image, and the pristine MoS$_2$ (2ML) spectrum is shown in black.
\textbf{b}, Constant height dI/dV maps ($V_\text{mod}$ = 20\,mV) at the sample voltages labelled in a. A progression of the orbital image is observed within certain resonances (cf. D1-D5) that might explain their large broadening. The resemblance of the highest occupied (D1) and lowest unoccupied defect orbital (C3) is indicative of a singly occupied defect state.
\textbf{c}, STM topography of neutral Re dopants (top layer) in bilayer MoS$_2$ on QFEG. The orientation of the distorted STM contrast along the crystal axes is indicated by colored arrows.
\textbf{d,e}, STM topography of Re$_\text{Mo}^{0}$ in bilayer MoS$_2$ in a different location also showing a preferential orientation. Here, the D$_\text{3h}$ symmetry breaking is observed at positive and negative bias, however the respective symmetry axis is canted by 60°.
}
\end{figure*}

Remarkably, we observe that the orbital image of the broad defect state resonance undergoes a continuous evolution of contrast, depending on the energy at which the defect state is probed, as shown in \figref{fig:STS_continuum_directional}a,b.
This behavior is observed on both mono- and bilayer MoS$_2$, although the direction of the progression varies depending on the precise location within the sample (cf. series \figref{fig:STS_continuum_directional}b and Fig.~S7c). This unusual behavior might be a consequence of the comparably flat adiabatic potential energy surface introduced by the pseudo-JTE~\cite{bersuker_jahn-teller_2017,bersuker_jahnteller_2021}, establishing a configurational continuum without significant energy barrier between the states (\figref{fig:theory}a-c). This phenomena is common for Jahn-Teller systems and is typically referred to as dynamic JTE~\cite{bersuker_jahn-teller_2017,bersuker_jahnteller_2021}. 
Such a configurational continuum may explain the unusual progression of the STS contrast within the defect resonance and its significant broadening.
\\

In most regions of the sample, the symmetry breaking direction is not a random distribution. Rather, there are larger domains where all Re$_\text{Mo}^{0}$ or Re$_\text{Mo}^{-1}$ relax in the same direction, as illustrated in \figref{fig:STS_continuum_directional}c-e. Directionally-aligned domains have been observed not only in highly doped (5\%) samples, but also in samples with comparably low doping concentrations ($<0.1$\%). 
Therefore, we suspect that residual strain in the epitaxially grown TMD, rather than direct dopant-dopant interactions, could be responsible for the aligned domains.
The flat APES may render the symmetry-broken Re particularly susceptible to external perturbations such as strain. In turn, this susceptibility may serve as a local strain sensor.
Further studies are needed to establish a firm correlation between the mesoscale strain tensor and local Jahn-Teller distortion of the individual Re dopants.
\\

\section*{Conclusions}
In summary, we have directly visualized the charge state-dependent symmetry breaking at S vacancies and Re substituting for Mo in mono- and bilayer MoS$_2$. Vac$_\text{S}^{-1}$ (in the top and bottom) coexists in both a threefold symmetric and symmetry-broken state. By controlling the Fermi level alignment through the substrate chemical potential, we are able to stabilize Re in three different charge states.
While Re$_\text{Mo}$ in the positive charge state retains the underlying D$_{3h}$ lattice symmetry, Re$_\text{Mo}^{0}$ and Re$_\text{Mo}^{-1}$ exhibit an increasingly distorted geometry.
We assign the minute local lattice relaxation detected by CO-tip nc-AFM and significant electronic reconfiguration of the defect states measured by STS to a pseudo-JTE, corroborated by DFT calculations.
A progressive defect orbital evolution observed within the unusually broad defect resonances is attributed to a flat potential energy surface, which leads to a continuum of the ground state geometry configuration.
Complementary measurements are needed to corroborate the link between the intriguing observation of directionally aligned Re domains and an inherent strain profile in the MoS$_2$ layers.
The charge-state dependent JTE exhibited by chalcogen vacancies and Re dopants offers a powerful means to precisely control the photo-physics of these prototypical atomic quantum emitters.

\section*{References}
\bibliographystyle{nature_bsc}
\bibliography{references.bib}

\clearpage

\section*{Figures}

\clearpage

\section*{Data availability}
The data that support the findings of this study are available from the corresponding author upon reasonable request.

\section*{Acknowledgements}
We thank Adam Gali for valuable discussions.
F.X. acknowledges the Walter Benjamin Programme from Deutsche Forschungsgemeinschaft (DFG). L.H., J.A. and B.S. appreciate funding from the European Research Council (ERC) under the European Union’s Horizon 2020 research and innovation program (Grant agreement No. 948243). 
R.T., Y.-C. L, and J.A.R. acknowledge funding from NEWLIMITS, a center in nCORE as part of the Semiconductor Research Corporation (SRC) program sponsored by NIST through award number 70NANB17H041 and the Department of Energy (DOE) through award number DESC0010697. R.T., C.D., and J.A.R. acknowledge funding from the 2D Crystal Consortium, National Science Foundation Materials Innovation Platform, under cooperative agreement DMR-1539916.
For the purpose of Open Access, the author has applied a CC BY public copyright license to any Author Accepted Manuscript version arising from this submission.

\section*{Author contributions}
\textsuperscript{\textdagger} These authors contribute equally to this work.

\section*{Competing interests}
The authors declare no competing interests.

\section*{Additional information}
\textbf{Supplementary information} The online version contains supplementary material
available at https://doi.org/\\

\textbf{Correspondence and requests for materials} should be addressed to Bruno Schuler.

\end{document}


\title{Supplemental Material:\texorpdfstring{\\}{}Charge State-Dependent Symmetry Breaking of Atomic Defects in Transition Metal Dichalcogenides}

\author{Feifei Xiang\,\orcidlink{0000-0002-4930-5919} \textsuperscript{\textdagger}}

\affiliation{nanotech@surfaces Laboratory, Empa -- Swiss Federal Laboratories for Materials Science and Technology, D\"ubendorf 8600, Switzerland}

\author{Lysander Huberich\textsuperscript{\textdagger}}
\affiliation{nanotech@surfaces Laboratory, Empa -- Swiss Federal Laboratories for Materials Science and Technology, D\"ubendorf 8600, Switzerland}

\author{Preston A. Vargas\textsuperscript{\textdagger}}
\affiliation{Department of Materials Science and Engineering, University of Florida, Gainesville, FL, 32611, USA}

\author{Riccardo Torsi}
\affiliation{Department of Materials Science and Engineering, The Pennsylvania State University, University Park, PA 16082, USA}

\author{Jonas Allerbeck\,\orcidlink{0000-0002-3912-3265}}
\affiliation{nanotech@surfaces Laboratory, Empa -- Swiss Federal Laboratories for Materials Science and Technology, D\"ubendorf 8600, Switzerland}

\author{Anne Marie Z. Tan}
\affiliation{Department of Materials Science and Engineering, University of Florida, Gainesville, FL, 32611, USA}
\affiliation{Institute of High Performance Computing (IHPC), Agency for Science, Technology and Research (A*STAR), Singapore 138632, Republic of Singapore}

\author{Chengye Dong}
\affiliation{Two-Dimensional Crystal Consortium, The Pennsylvania State University, University Park, PA 16802, USA}

\author{Pascal Ruffieux\,\orcidlink{0000-0001-5729-5354}}
\affiliation{nanotech@surfaces Laboratory, Empa -- Swiss Federal Laboratories for Materials Science and Technology, D\"ubendorf 8600, Switzerland}

\author{Roman Fasel\,\orcidlink{0000-0002-1553-6487}}
\affiliation{nanotech@surfaces Laboratory, Empa -- Swiss Federal Laboratories for Materials Science and Technology, D\"ubendorf 8600, Switzerland}

\author{Oliver Gr\"oning}
\affiliation{nanotech@surfaces Laboratory, Empa -- Swiss Federal Laboratories for Materials Science and Technology, D\"ubendorf 8600, Switzerland}

\author{Yu-Chuan Lin}
\affiliation{Department of Materials Science and Engineering, The Pennsylvania State University, University Park, PA 16082, USA}
\affiliation{Department of Materials Science and Engineering, National Yang Ming Chiao Tung University, Hsinchu City 300, Taiwan}

\author{Richard G. Hennig\,\orcidlink{0000-0003-4933-7686}}
\affiliation{Department of Materials Science and Engineering, University of Florida, Gainesville, FL, 32611, USA}

\author{Joshua A. Robinson\,\orcidlink{0000-0002-1513-7187}}
\affiliation{Department of Materials Science and Engineering, The Pennsylvania State University, University Park, PA 16082, USA}
\affiliation{Two-Dimensional Crystal Consortium, The Pennsylvania State University, University Park, PA 16802, USA}
\affiliation{Department of Chemistry and Department of Physics, The Pennsylvania State University, University Park, PA, 16802, USA}

\author{Bruno Schuler\,\orcidlink{0000-0002-9641-0340}}
\email[]{bruno.schuler@empa.ch}
\affiliation{nanotech@surfaces Laboratory, Empa -- Swiss Federal Laboratories for Materials Science and Technology, D\"ubendorf 8600, Switzerland}

\date{\today}
\pacs{}
\maketitle

\tableofcontents
\newpage

\clearpage

\section{Methods}

\subsection*{Sample preparation}

Metal-organic chemcial vapor deposition was used to grow monolayer MoS$_2$ on an epitaxial graphene and quasi-freestanding epitaxial graphene on 6H-SiC(0001) substrate H$_2$S and Mo(CO)$_6$ were used as precursors for growth of MoS$_2$, while the Re dopants were introduced by admixing 0.1$\%$, 5$\%$ of Re$_2$(CO)$_{10}$ during the growth of MoS$_2$. Before SPM inspection, all samples were annealed to 300$^\circ$C for about 30 min to remove the possible adsorbates. S vacancies in MoS$_2$ were created by annealing the pristine MoS$_2$ to 900$^\circ$C for 30 min. 
Epitaxial graphene was synthesized \textit{via} silicon sublimation from the Si face of SiC~\cite{emtsev2009towards}, subsequently monolayer epitaxial graphene is annealed at 950$^\circ$C for 30 min in pure hydrogen to intercalate hydrogen between the buffer layer and silicon carbide, yielding bilayer quasi-freestanding epitaxial graphene~\cite{robinson2011epitaxial}.

\subsection*{Scanning probe microscopy (SPM) measurements}
SPM measurements were acquired with a Scienta-Omicron GmbH or CreaTec Fischer \& Co. GmbH scanning probe microscope at liquid helium temperatures ($T <$ 5\,K) under ultrahigh vacuum ($p<$ 2$\times$10$^{-10}$\,mbar). The quartz crystal cantilever (qPlus based) sensor~\cite{giessibl_high-speed_1998} ($f_0 \approx 27\,$kHz, $Q \approx 47\,k$) tip apex was prepared by indentations into a gold substrate and verified as metallic on a Au(111) surface. Non-contact AFM images were taken with a carbon monoxide functionalized tip~\cite{mohn_different_2013} in constant height mode at zero bias with an oscillation amplitude $A_{\text{osc}} < 100$\,pm. STM topographic measurements were taken in constant current feedback with the bias voltage applied to the sample. Scanning tunneling spectroscopic (STS) measurements were recorded using a lock-in amplifier (HF2LI Lock-in Amplifier from Zurich Instruments or Nanonis Specs) with a resonance frequency between 600-700\,Hz and a modulation amplitude provided in the figure caption.

\subsection{First-principles calculations}
We used density functional theory (DFT) as implemented in the Vienna \textit{ab initio} simulation package VASP\cite{Kresse1996a} to evaluate the projected densities of states. We modeled the core electrons using projector-augmented wave (PAW) potentials \cite{Blochl1994,Joubert1999} with valence electron configurations of $4p^64d^55s^1$ and $3s^23p^4$ for Mo and S, respectively. We performed spin-polarized calculations with a plane wave cutoff energy of 520\,eV, which ensures energy convergence to within 1\,meV/atom with $\Gamma$-centered Monkhorst-Pack k-point meshes\cite{Monkhorst1976} for Brillouin zone integration. In addition, we enforced integer band occupancies during our calculations to avoid unphysical electronic ground states with fractional band occupancies, which may arise due to wavefunction overlap.\\

In the sulfur vacancy calculations, we used 4$\times$4 supercells constaining a single vacancy each, and 20\,Å vacuum spacing between layers. We also treated the exchange-correlation using the strongly constrained and appropriately normed meta-generalized gradient approximation functional with van der Waals interactions (SCAN+rVV10)\cite{Peng2016}. For the projected density of states calculations, we used Gaussian smearning with a reduced smearing width of 0.02\,eV and a 6$\times$6$\times$2 k-mesh.\\

In the Re substitution calculations, we used 5$\times$5 supercells containing a single substitution each, and 20\,Å vacuum spacing between layers. We treated the exchange-correlation using the PBE functional\cite{Perdew1996} with VdW contributions being accounted for by the DFTD3 method of Grimme et. al.\cite{Grimme2010}. For the projected density of states, we used the tetrahedron method with a 5$\times$5$\times$1 k-point mesh.

\clearpage

\section{Supplementary Experiments}

\begin{figure*}[h!]
\includegraphics[width=\textwidth]{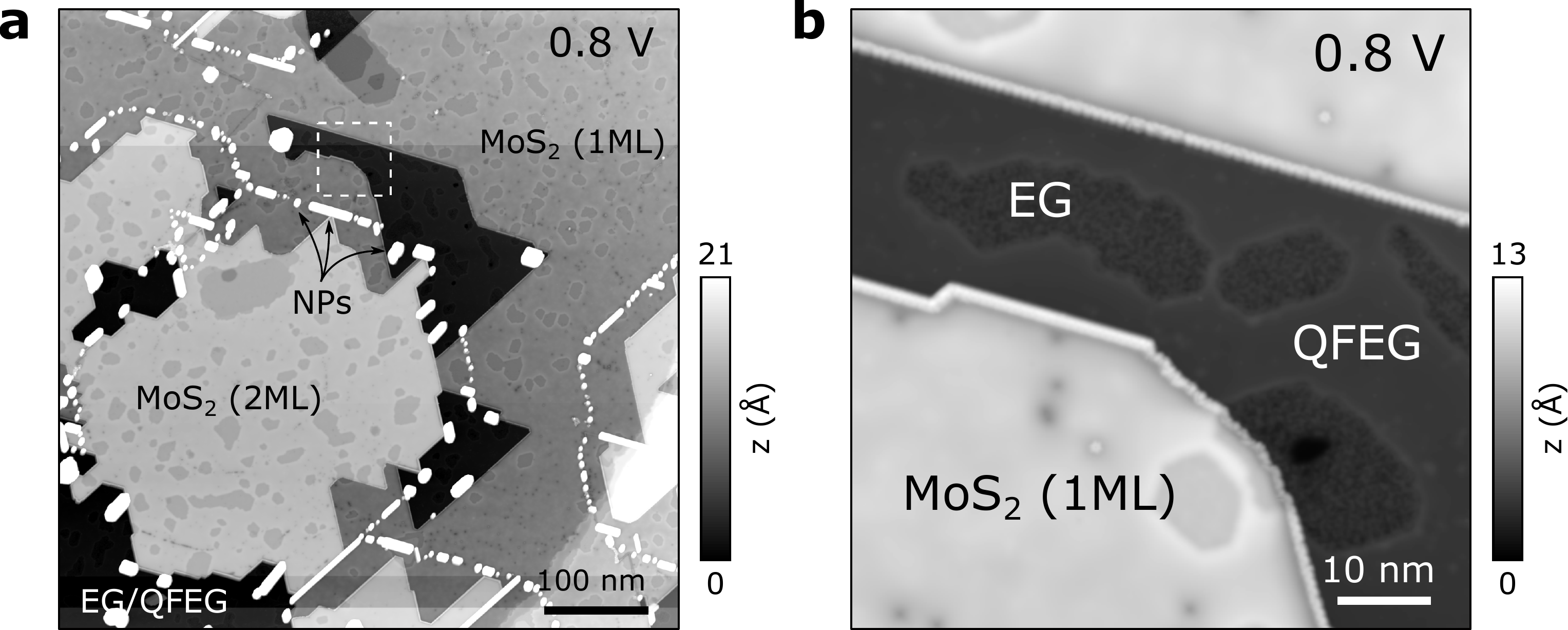}
\caption{\label{fig:overview_pristine}
\textbf{Monolayer and bilayer MoS$_2$ on EG and QFEG.}
\textbf{a}, Large-scale (800\,nm $\times$ 800\,nm) STM topography ($I = 50\,$pA) of mono- and bilayer MoS$_2$ on mostly hydrogen-intercalated epitaxial graphene. \textbf{b}, Close-up of dashed rectangle in a, showing contrast difference between EG and QFEG areas, and how they appear if overgrown by MoS$_2$.}
\end{figure*}

\clearpage

\begin{figure*}[h!]
\includegraphics[width=\textwidth]{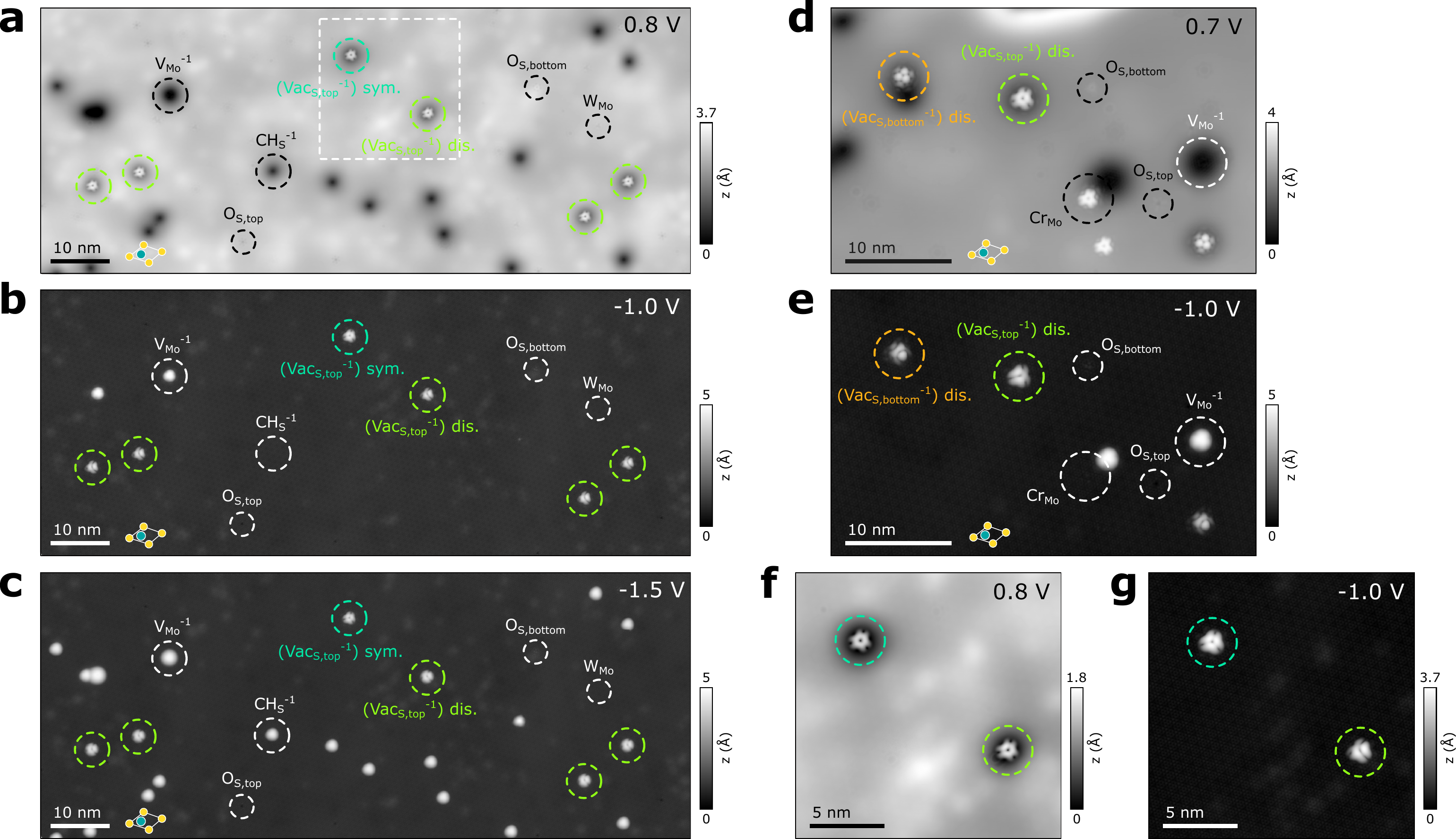}
\caption{\label{fig:VacS_overview}
\textbf{Monolayer MoS$_2$/QFEG defects overview.}
\textbf{a-g}, STM topographies ($I = 50\,$pA) of annealing-induced negatively charged sulfur vacancies (Vac$_\text{S}^{-1}$) (colored circles) in  monolayer MoS$_2$/QFEG along with other as-grown defects such as O$_\text{S}$~\cite{barja_identifying_2019,schuler_how_2019}, CH$_\text{S}$~\cite{cochrane_intentional_2020}, V$_\text{Mo}$~\cite{kozhakhmetov_controllable_2021}, Cr$_\text{Mo}$~\cite{schuler_how_2019}, and W$_\text{Mo}$, labelled in black. 
Top and bottom Vac$_\text{S}^{-1}$ can be distinguished by their distinct STM contrast~\cite{schuler_large_2019,mitterreiter_atomistic_2020}.
The symmetric and distorted Vac$_\text{S}^{-1}$ can be identified at negative bias voltage. A close-up of the dashed box in a is shown in f,g.}
\end{figure*}

\clearpage

\begin{figure*}[h!]
\includegraphics[width=0.7\textwidth]{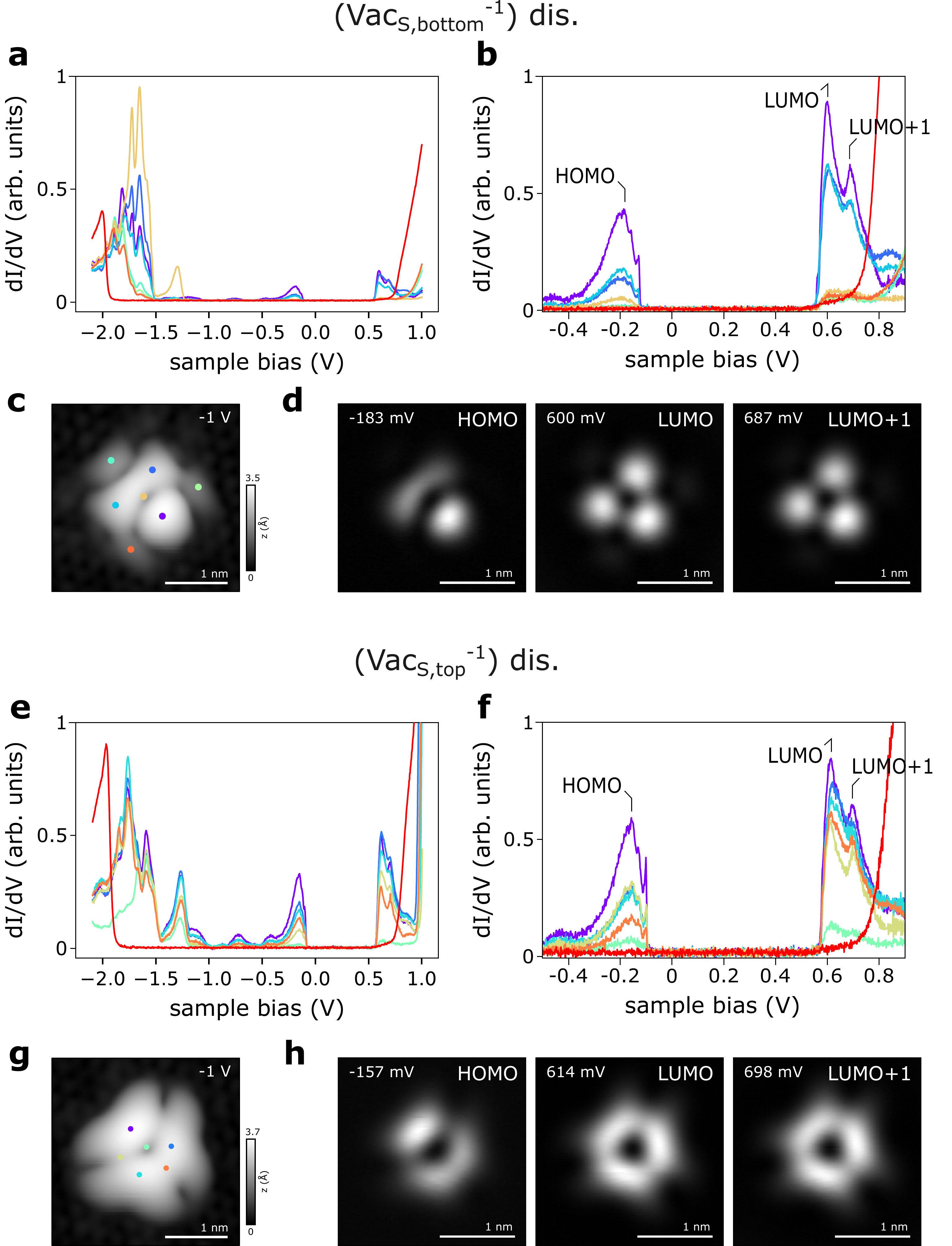}
\caption{\label{fig:VacS_STS_lr}
\textbf{STS of distorted Vac$_\text{S,top}^{-1}$ and Vac$_\text{S,bottom}^{-1}$.}
\textbf{a,b}, d$I$/d$V$ spectra of a distorted Vac$_\text{S,bottom}^{-1}$ at the locations indicated in c.  The red spectrum is taken on the pristine MoS$_2$ surface.
\textbf{c}, STM image of Vac$_\text{S,bottom}^{-1}$. 
\textbf{d}, d$I$/d$V$ maps of the frontier orbitals labelled in b.
\textbf{e,f}, d$I$/d$V$ spectra of a distorted Vac$_\text{S,top}^{-1}$ at the locations indicated in g. The red spectrum is taken on the pristine MoS$_2$ surface.
\textbf{g}, STM image of Vac$_\text{S,top}^{-1}$.  
\textbf{h}, d$I$/d$V$ maps of the frontier orbitals labelled in f.
The HOMO orbitals for both the bottom and top S vacancy resembles the same asymmetry as observed in the corresponding STM topography at negative bias. Note that the asymmetry of the Vac$_\text{S,top}^{-1}$ HOMO shown in h is different from the one shown in Fig.~1e. Both orbital shapes are regularly observed and may indicate a different local strain profile as discussed in the main text.
$V_\text{mod}$ = 10\,mV (a,e), 2\,mV (b,f), and 20\,mV (d,h).}
\end{figure*}

More detailed conductance measurements of Vac$_\text{S}^{-1}$ are shown in \figref{fig:VacS_STS_lr}. We measured the conductance (d$I$/d$V$) at several positions on each defect, the corresponding dI/dV spectra are shown in \figref{fig:VacS_STS_lr}a, b, e, f, respectively. Apart from the frontier orbitals around zero bias, several hydrogenic bound states due to the negative charge can be identified above the valence band edge~\cite{aghajanian_resonant_2020} (\figref{fig:VacS_STS_lr}a and e). If we just focus on the frontier orbitals of both the distorted Vac$_\text{S,bottom}^{-1}$ and Vac$_\text{S,top}^{-1}$ (\figref{fig:VacS_STS_lr}b and f), three resonances are revealed with with several vibronic excitations visible. 
As seen in the constant height dI/dV maps in \figref{fig:VacS_STS_lr}d and h, the highest occupied defect resonance (HOMO) exhibits a clear two-fold symmetric shape, whereas the two unoccupied orbitals appear basically three-fold symmetric.

\clearpage

\begin{figure*}[h!]
\includegraphics[width=\textwidth]{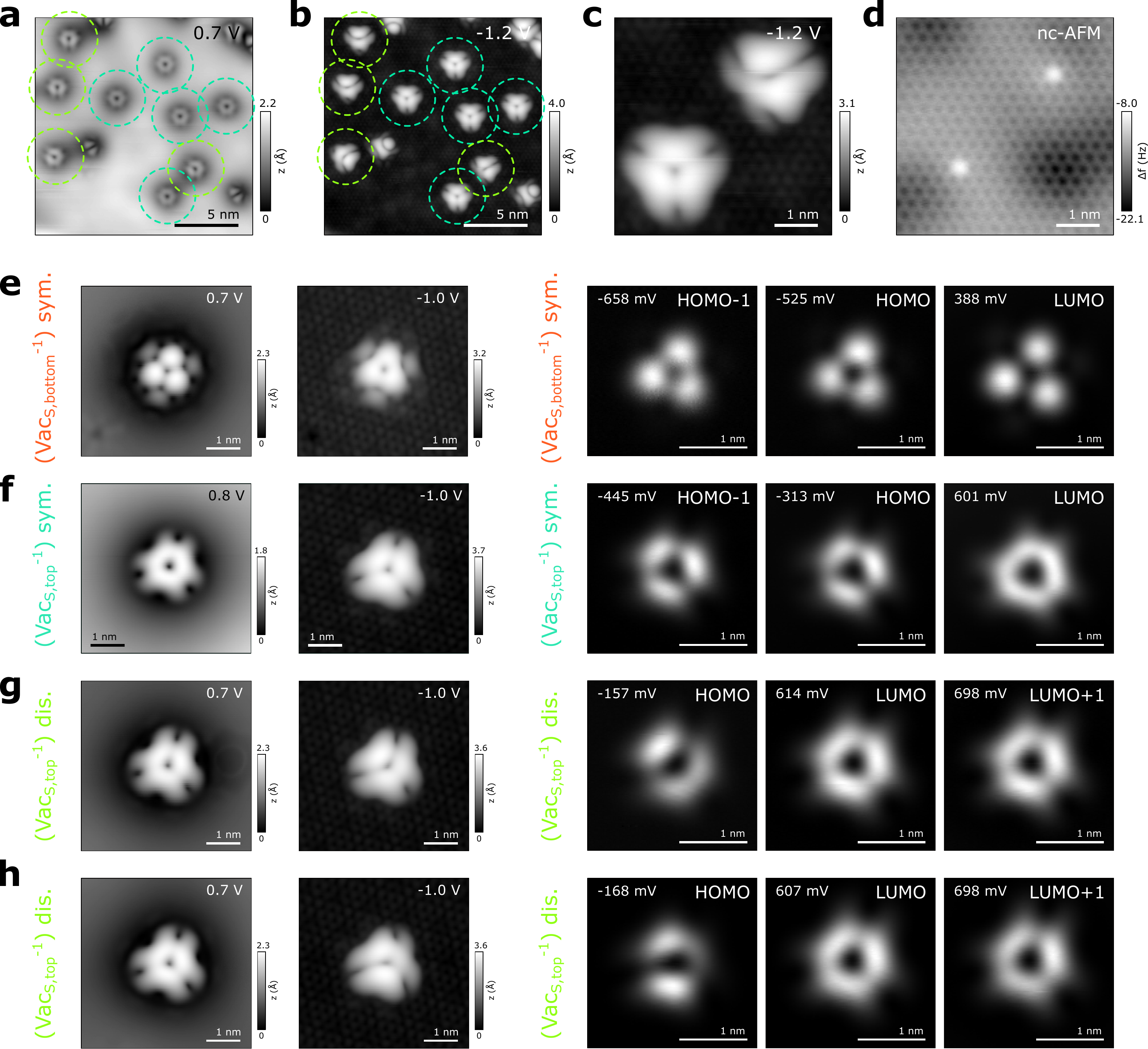}
\caption{\label{fig:VacS_other}
\textbf{Symmetric and distorted Vac$_\text{S}^{-1}$ comparison.}
\textbf{a,b}, STM topography ($I = 50\,$pA) of negatively charged sulfur top vacancies in MoS$_2$/QFEG at positive (a) and negative (b) sample bias. 
\textbf{c,d}, STM topography ($I = 50\,$pA) and nc-AFM image (metal tip) of a symmetric and distorted Vac$_\text{S, top}^{-1}$.
\textbf{e-h}, STM topographies ($I = 50\,$pA) (left) and d$I$/d$V$ maps (right) of the symmetric Vac$_\text{S, bottom}^{-1}$ (e), symmetric Vac$_\text{S, top}^{-1}$ (f), and two versions of a distorted Vac$_\text{S, top}^{-1}$ (g,h) for comparison. Panels c,d and h are the same as in Fig.~1b,c, and Fig.~1e, but are reprinted for comparison.}
\end{figure*}

At positive bias, all Vac$_\text{S, top}^{-1}$ appear threefold symmetric, while at negative bias some appear threefold symmetric (turquoise circles in \figref{fig:VacS_overview} and \figref{fig:VacS_other}b), while other appear at reduced twofold symmetry (light green circle) at about an equal share. The high resolution STM images of symmetric and distorted Vac$_\text{S}^{-1}$ and the corresponding d$I$/d$V$ maps taken in constant height mode are displayed in \figref{fig:VacS_other}e-h. The orbital symmetry breaking is most clearly revealed in the defect's HOMO orbital (cf.~\figref{fig:VacS_other}g,h).

\clearpage

\begin{figure*}[h!]
\includegraphics[width=\textwidth]{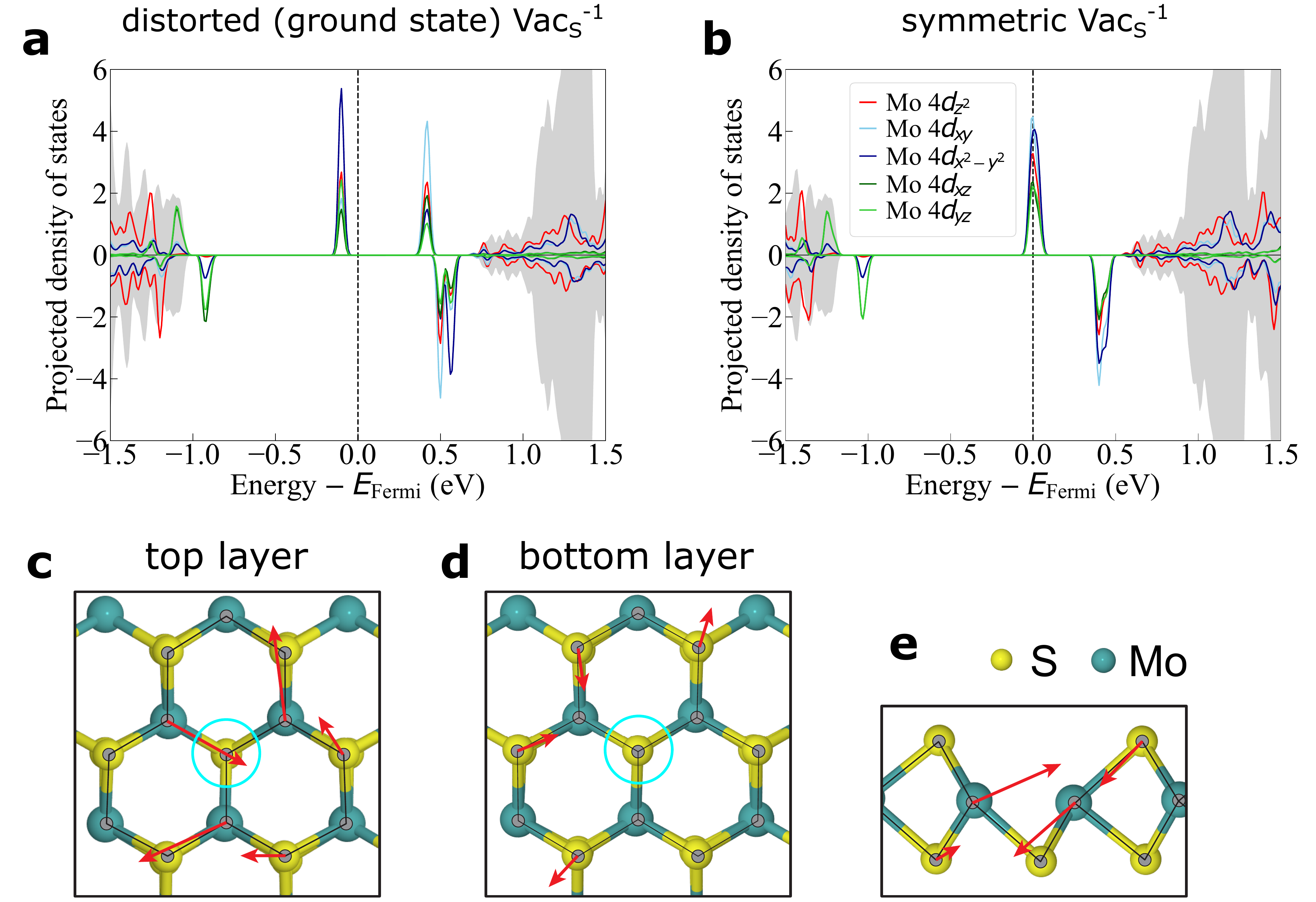}
\caption{\label{fig:VacS_theory}
\textbf{Projected density of states of the distorted and symmetric Vac$_\text{S}^{-1}$ in MoS$_2$.}
\textbf{a}, Vac$_\text{S}^{-1}$ in the constrained symmetric geometry ($+116$\,meV~\cite{tan_stability_2020}) calculated in a 4$\times$4 supercell. 
\textbf{b}, Vac$_\text{S}^{-1}$ in the distorted ground state geometry calculated in a 4$\times$4 supercell. The density of states is projected onto the Mo $d$ orbitals. 
\textbf{c-e}, calculated geometry of MoS$_2$ with a distorted Vac$_\text{S}^{-1}$, looking from the top layer (c), bottom layer (d) and side (e). The red arrows show the atom relaxations compared to the MoS$_2$ lattice with a symmetric Vac$_\text{S}^{-1}$ (atom positions are shown as gray ball and stick model). The position of Vac$_\text{S}^{-1}$ is located on the top layer of MoS$_2$ and is marked by cyan circle. The arrow length is 50 times longer than the actual relaxation distance.
}
\end{figure*}

The DFT calculated projected density of states (pDOS) of symmetric and distorted Vac$_\text{S}^{-1}$ are shown in \figref{fig:VacS_theory}a and b. In the symmetric Vac$_\text{S}^{-1}$, a single electron is occupying the two degenerate orbitals at the Fermi level, thereby generating a Jahn-Teller instability. The degenerate orbitals have a comparable contribution from the $d_{x^2-y^2}$ (dark blue) and $d_{xy}$ (light blue) Mo orbital. In the distorted configuration, this degeneracy is lifted such that only a non-degenerate, fully occupied state exists with a predominant $d_{x^2-y^2}$ character (dark blue). The local JT displacements in the distorted geometry are shown in \figref{fig:VacS_theory}c-e. Mo atoms adjacent to the Vac$_\text{S}$ exhibit the largest displacemebts on the order of 5\,pm compared to the symmetric case.\\

\clearpage

\begin{figure*}[h!]
\includegraphics[height=7cm]{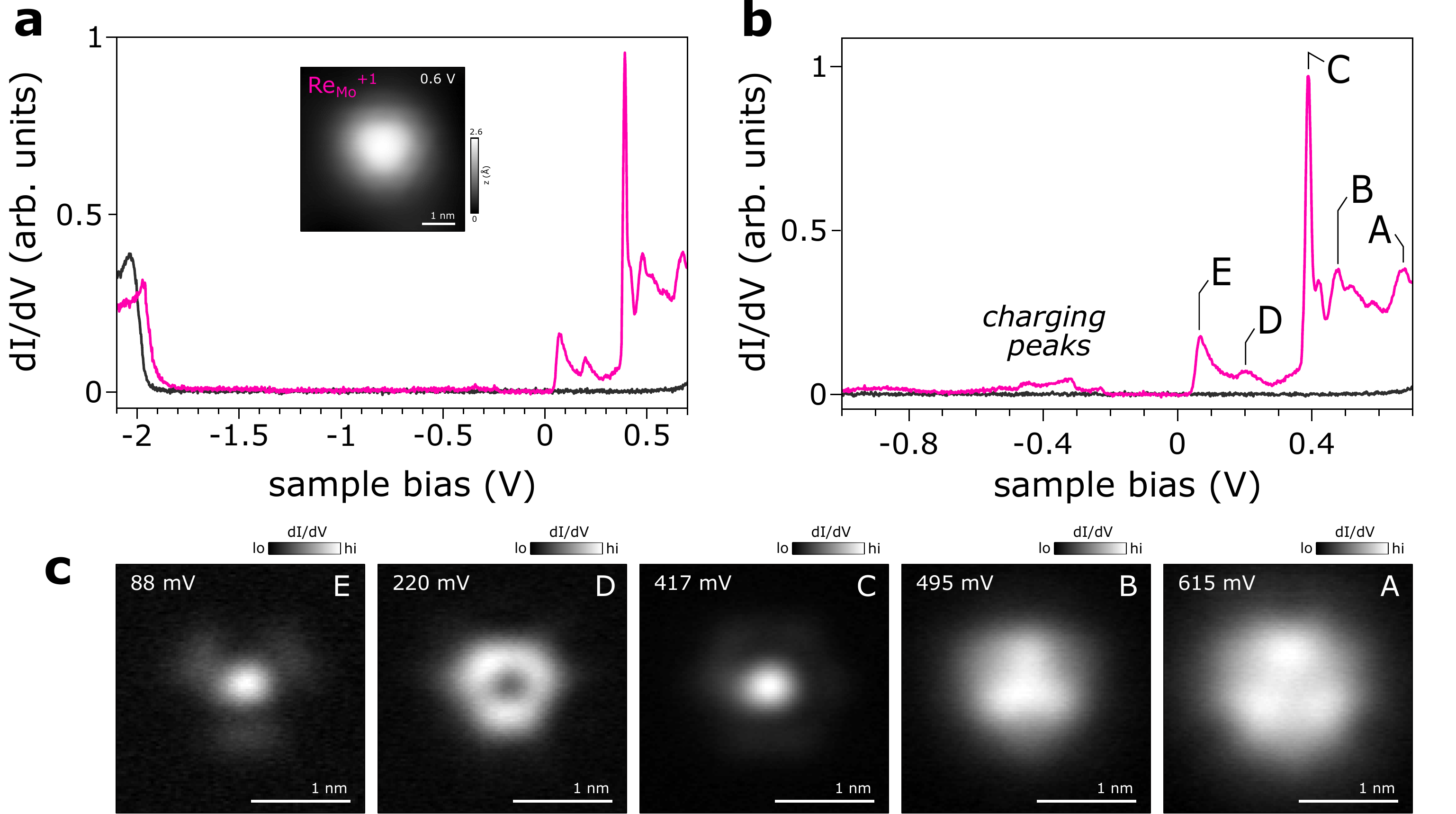}
\caption{\label{fig:Re_Mo_dIdV_+1}
\textbf{dI/dV spectra and orbital imaging of Re$_\text{Mo}^{+1}$ on monolayer MoS$_2$/QFEG.}
\textbf{a},\textbf{b} dI/dV spectra of Re$_\text{Mo}^{+1}$ with the main defect resonances indicated.
\textbf{c}, Constant height dI/dV maps of the defect orbitals labelled in a and b. Lock-in modulation: $V_\text{mod}$ = 20\,mV (a and c) and $V_\text{mod}$ = 10\,mV (b).
}
\end{figure*}

\clearpage

\begin{figure*}[h!]
\includegraphics[height=7cm]{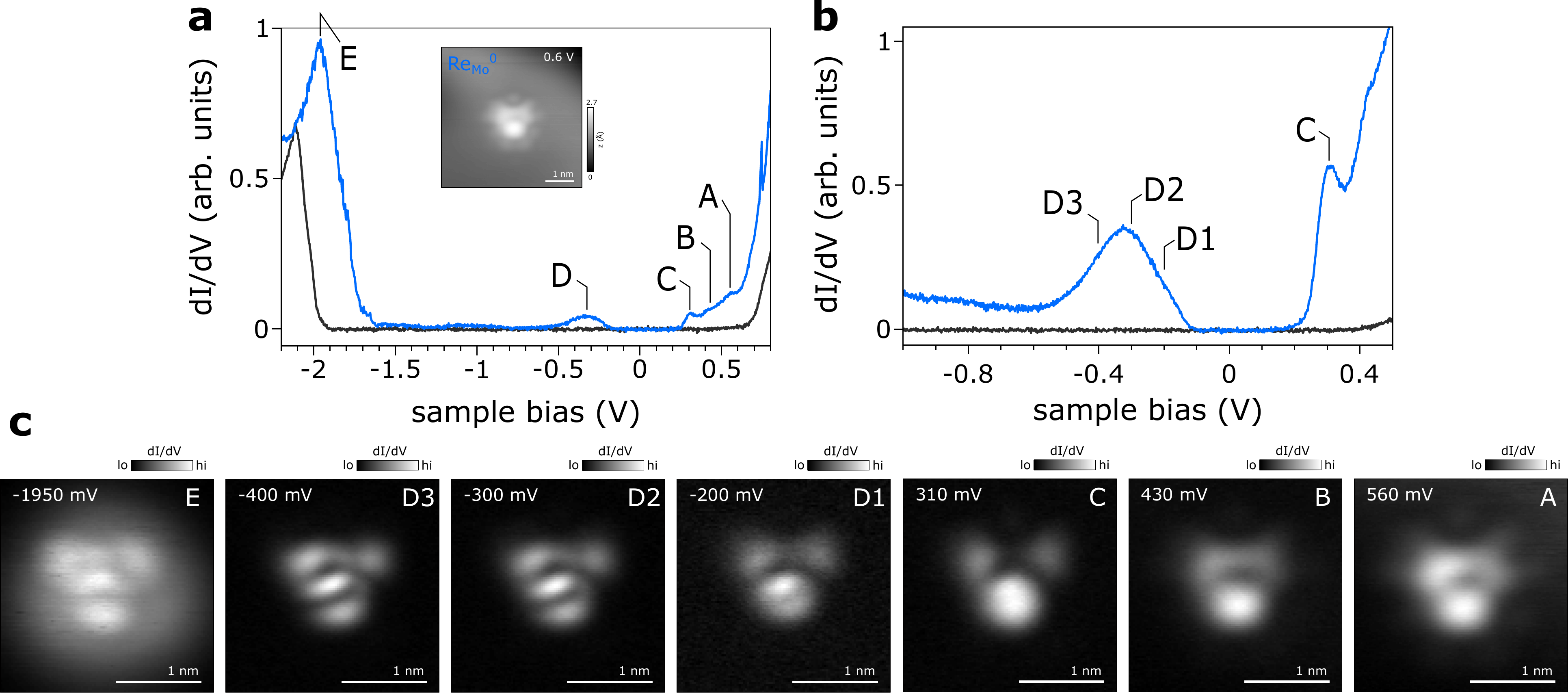}
\caption{\label{fig:Re_Mo_dIdV_0}
\textbf{dI/dV spectra and orbital imaging of Re$_\text{Mo}^{0}$ on monolayer MoS$_2$/QFEG.}
\textbf{a},\textbf{b} dI/dV spectra of Re$_\text{Mo}^{0}$ with the main defect resonances indicated.
\textbf{c}, Constant height dI/dV maps of the defect orbitals labelled in a and b. Note that the defect state evolution of the resonance D proceeds along a different crystallographic direction as compared to Fig.~5 in the main text. Lock-in modulation: $V_\text{mod}$ = 20\,mV (a and c) and $V_\text{mod}$ = 10\,mV (b)
}
\end{figure*}

\clearpage

\begin{figure*}[h!]
\includegraphics[height=7cm]{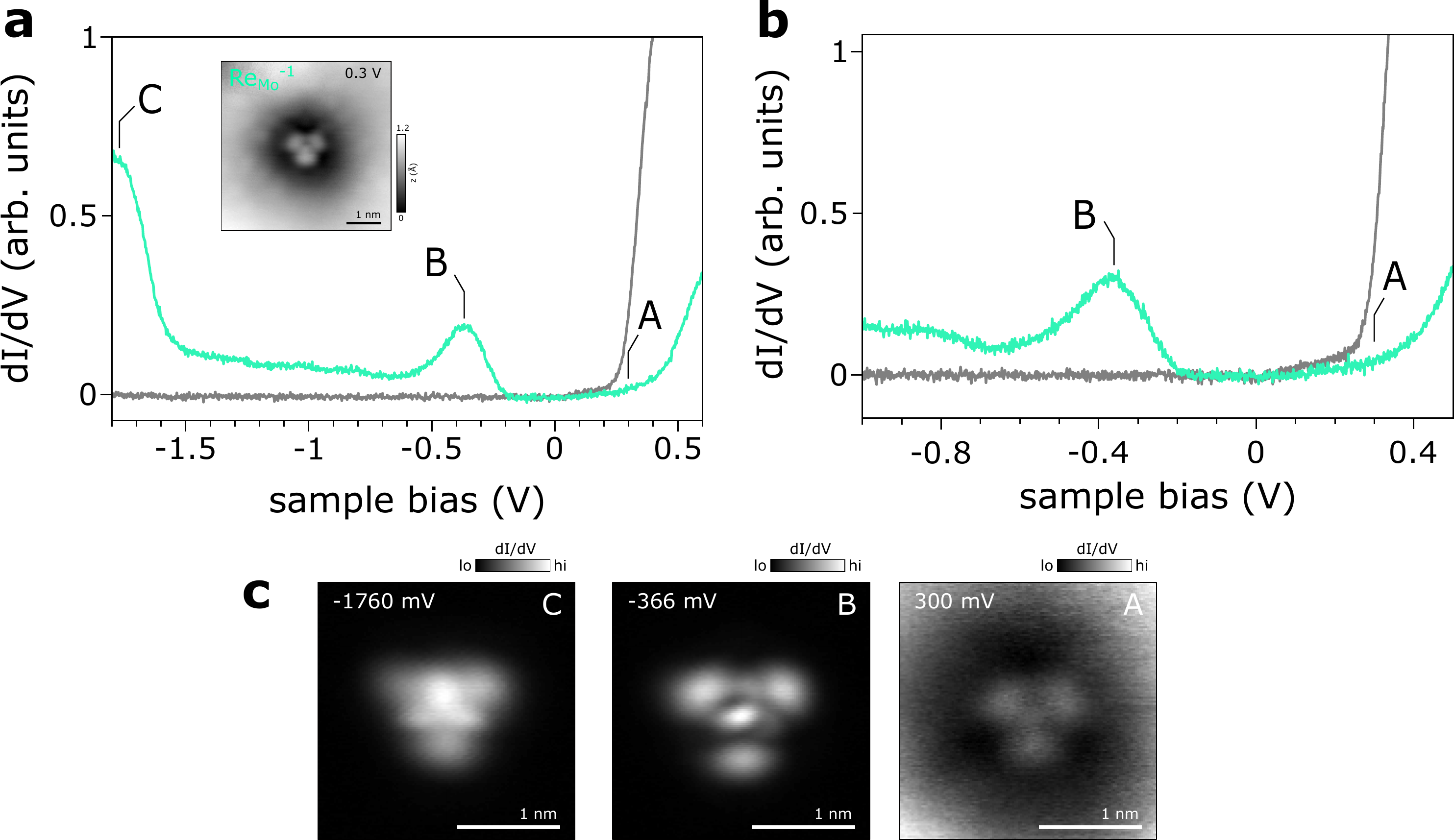}
\caption{\label{fig:Re_Mo_dIdV_-1}
\textbf{dI/dV spectra and orbital imaging of Re$_\text{Mo}^{-1}$ on monolayer MoS$_2$/EG.}
\textbf{a},\textbf{b} dI/dV spectra of Re$_\text{Mo}^{-1}$ with the main defect resonances indicated.
\textbf{c}, Constant height dI/dV maps of the defect orbitals labelled in a and b.
Note that the defect state evolution as discussed in the main text is not shown here. Lock-in modulation: $V_\text{mod}$ = 20\,mV (a and c) and $V_\text{mod}$ = 10\,mV (b)
}
\end{figure*}

\clearpage

\begin{figure*}[h!]
\includegraphics[width=\textwidth]{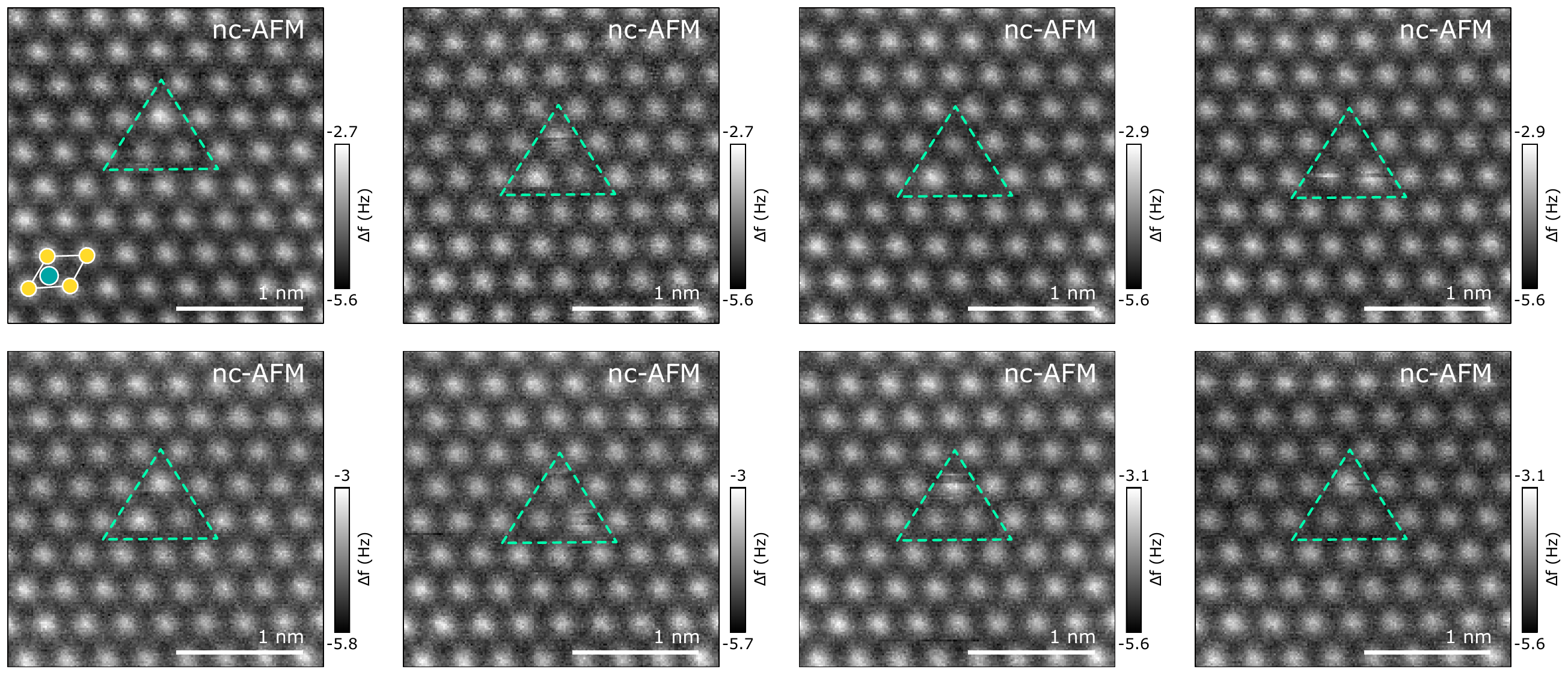}
\caption{\label{fig:switching}
\textbf{Conformational switching of Re$_\text{Mo}^{-1}$.} Consecutive CO-tip nc-AFM images of the same Re$_\text{Mo}^{-1}$ impurity. Top sulfur atoms are repulsive (bright) and molybdenum atoms attractive (dark). The MoS$_2$ unit cell is indicated at the bottom left of the first image and the position of the Re impurity is highlighted by the green triangle. One of the three neighboring top S atoms next to each Re impurity is protruding out of plane (brightest atom). Within or in-between consecutive images the protruding S atom is switched between one of the three equivalent lattice sites.
}
\end{figure*}

\clearpage

\begin{figure*}[h!]
\includegraphics[width=0.7\textwidth]{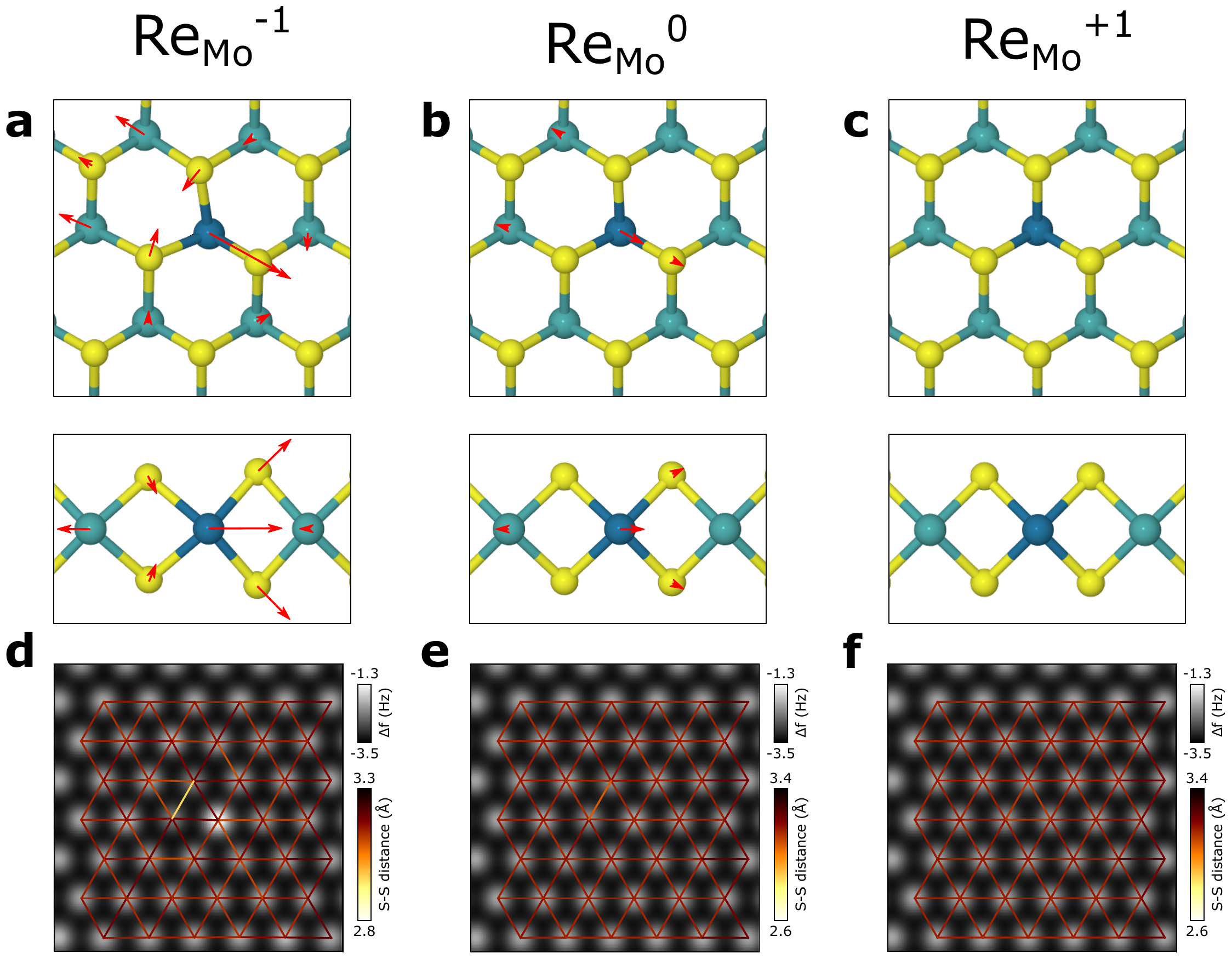}
\caption{\label{fig:strain_theo}
\textbf{Calculated geometry of Re$_\text{Mo}$ in different charge states.}
\textbf{a-c}, Ground state geometry of Re$_\text{Mo}$ in the -1\,$e$ (a), 0\,$e$ (b), and +1\,$e$ (c) charge state, respectively. The red arrows depict the atom relaxation as compared to the pristine MoS$_2$ lattice. The arrows are ten times longer than the actual relaxation. Both Re$_\text{Mo}^{-1}$ and Re$_\text{Mo}^{0}$ exhibit considerable Jahn-Teller distortions, while Re$_\text{Mo}^{+1}$ preserves a symmetric lattice geometry. \textbf{d-f}, Simulated CO-tip nc-AFM images of the defect geometries in a-c. The \df contrast with one protruding S atom, adjacent to the Re impurity for Re$_\text{Mo}^{-1}$ and to a lesser degree for Re$_\text{Mo}^{0}$ is in excellent agreement with experiment (see Fig.~3). The apparent sulfur-sulfur distances in the top S layer is indicated by the red-to-yellow lines. Significant lateral compressive strain introduced by the Re$_\text{Mo}^{-1}$ impurity is seen at nearest-neighbor S atoms opposite to the protruding S atom.  
}
\end{figure*}

\clearpage

\begin{figure*}[h!]
\includegraphics[width=\textwidth]{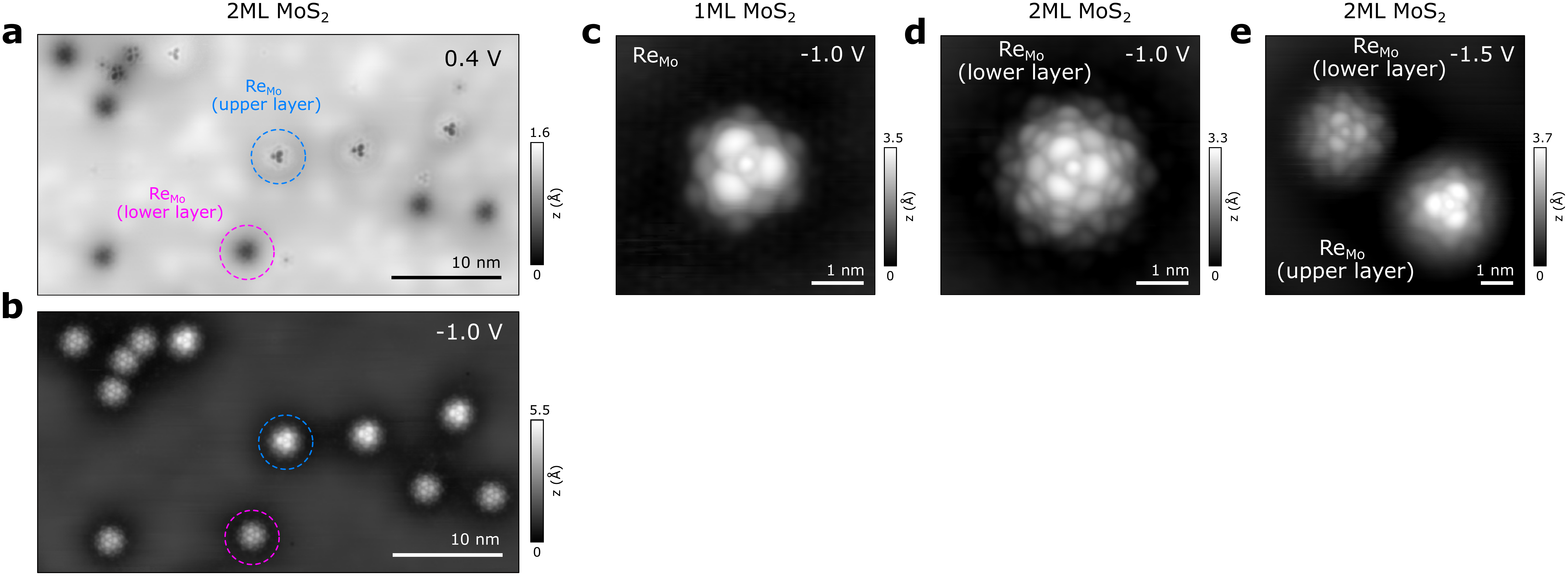}
\caption{\label{fig:multilayer_STM}
\textbf{STM contrast of Re$_\text{Mo}$ on bilayer MoS$_2$.} 
\textbf{a, b}, STM topography ($I = 50\,$pA) of Re impurities in bilayer MoS$_2$ on EG scanned at 0.4\,V and -1\,V, respectively. 
\textbf{c, d}, STM orbital image of Re$_\text{Mo}$ in monolayer and bilayer MoS$_2$. The Re$_\text{Mo}$ in d is located in the lower MoS$_2$ layer. \textbf{e} STM orbital image of two Re$_\text{Mo}$ in bilayer MoS$_2$. The upper left  impurity is located in the lower MoS$_2$ layer and the lower right impurity in the upper MoS$_2$ layer, with a brighter and mirrored appearance, as expected for 2H stacking.
}
\end{figure*}

Re$_\text{Mo}$ on bilayer MoS$_2$ on EG/SiC is shown in \figref{fig:multilayer_STM}. 
Two types of Re$_\text{Mo}$ defects can be found on the surface: Re$_\text{Mo}$ in the lower layer MoS$_2$ (marked by pink circle in \figref{fig:multilayer_STM}a and b) and Re$_\text{Mo}$ in the upper layer (marked by blue circle in \figref{fig:multilayer_STM}a and b). Strikingly, the the upper and lower layer Re$_\text{Mo}$ in 2ML MoS$_2$ exhibits an interference pattern around the defect, which is not seen for monolayer MoS$_2$ (cf. \figref{fig:multilayer_STM}c and d). The interference pattern may originate from quasi-particle scattering of dispersive MoS$_2$ bilayer states. In bilayer MoS$_2$, the Re$_\text{Mo}$ of the lower and upper layer are mirrored as expected for the 2H stacking (see \figref{fig:multilayer_STM}e).

\clearpage

\begin{figure*}[h!]
\includegraphics[width=0.8\textwidth]{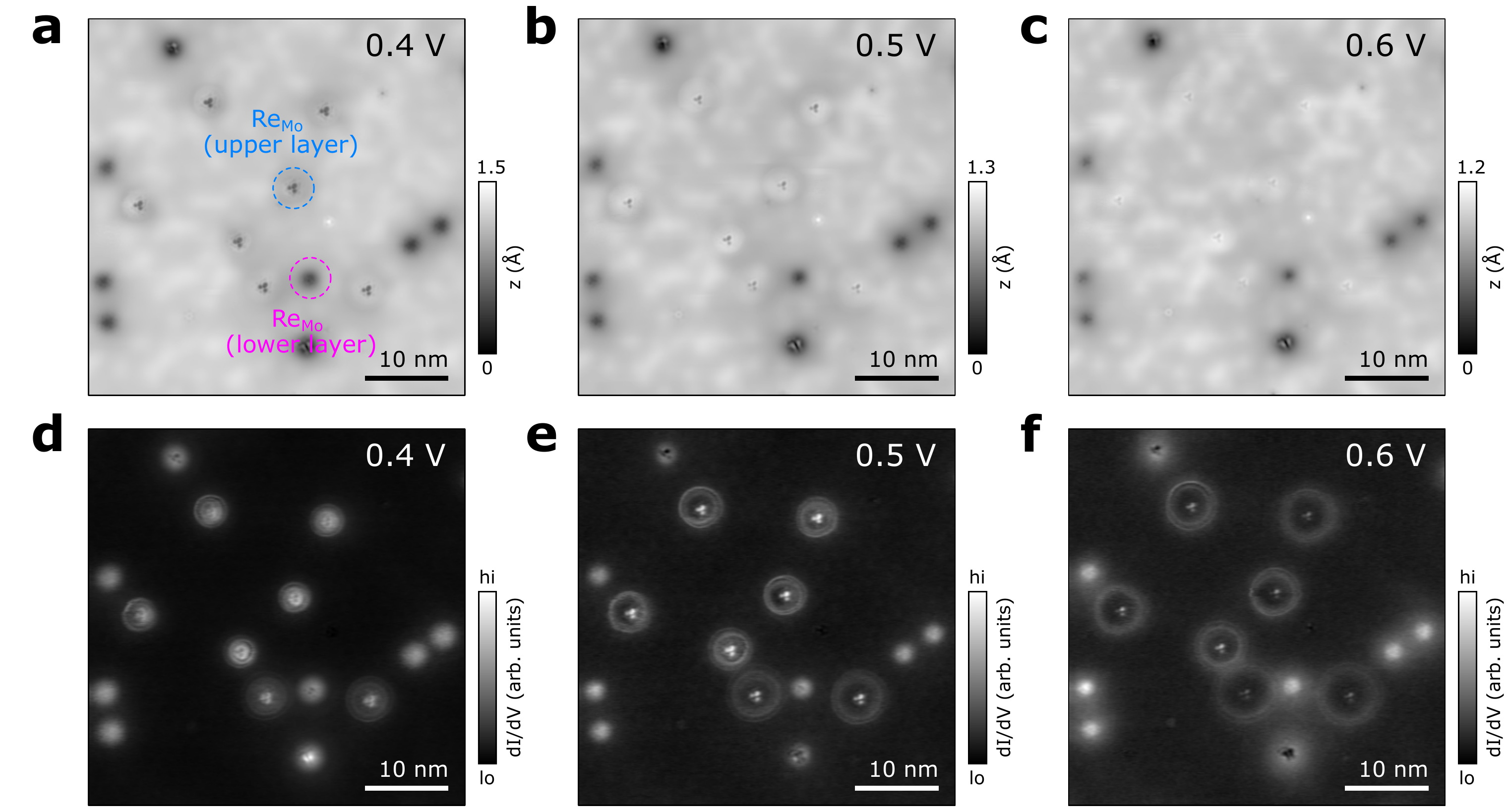}
\caption{\label{fig:chargingring_STM}
\textbf{Tip-induced charging rings at Re impurities in bilayer MoS$_2$.} 
\textbf{a-c}, STM topographies ($I = 50\,$pA) of Re impurities in the upper and lower MoS$_2$ layer grown on EG imaged at 0.4\,V (a), 0.5\,V (b), and 0.6\,V (c), respectively.  
\textbf{d-f}, d$I$/d$V$ channel recorded simultaneously with the STM topographies a-c ($V_\text{mod} = 20$\,mV). 
}
\end{figure*}

At positive sample bias, charging rings around the Re impurities that are induced by the tip electric field could be observed (\figref{fig:chargingring_STM}). Such rings only appear for Re$_\text{Mo}$ in the upper MoS$_2$ layer. 
Re impurities in the lower layer appear as dark protrusions similar to  on 1ML MoS$_2$/EG. Re impurities in the upper layer, however, have a dark tri-lobal contrast surrounded by two concentric bright ring (\figref{fig:chargingring_STM}a-c). These concentric rings can be best seen in d$I$/d$V$ maps recorded in constant current mode, shown in \figref{fig:chargingring_STM}d-f. As expected the ring size increases with increasing bias. Presumably, the Re$_\text{Mo}$ in the upper layer are initially charge neutral but a small tip-electric field can already decharge them. Inside the charging rings, the defect is therefore negatively charged and the dark contrast corresponds to the depopulated defect orbital.

\clearpage

\section{Supplementary Calculations}

\begin{figure*}[h!]
\includegraphics[width=\textwidth]{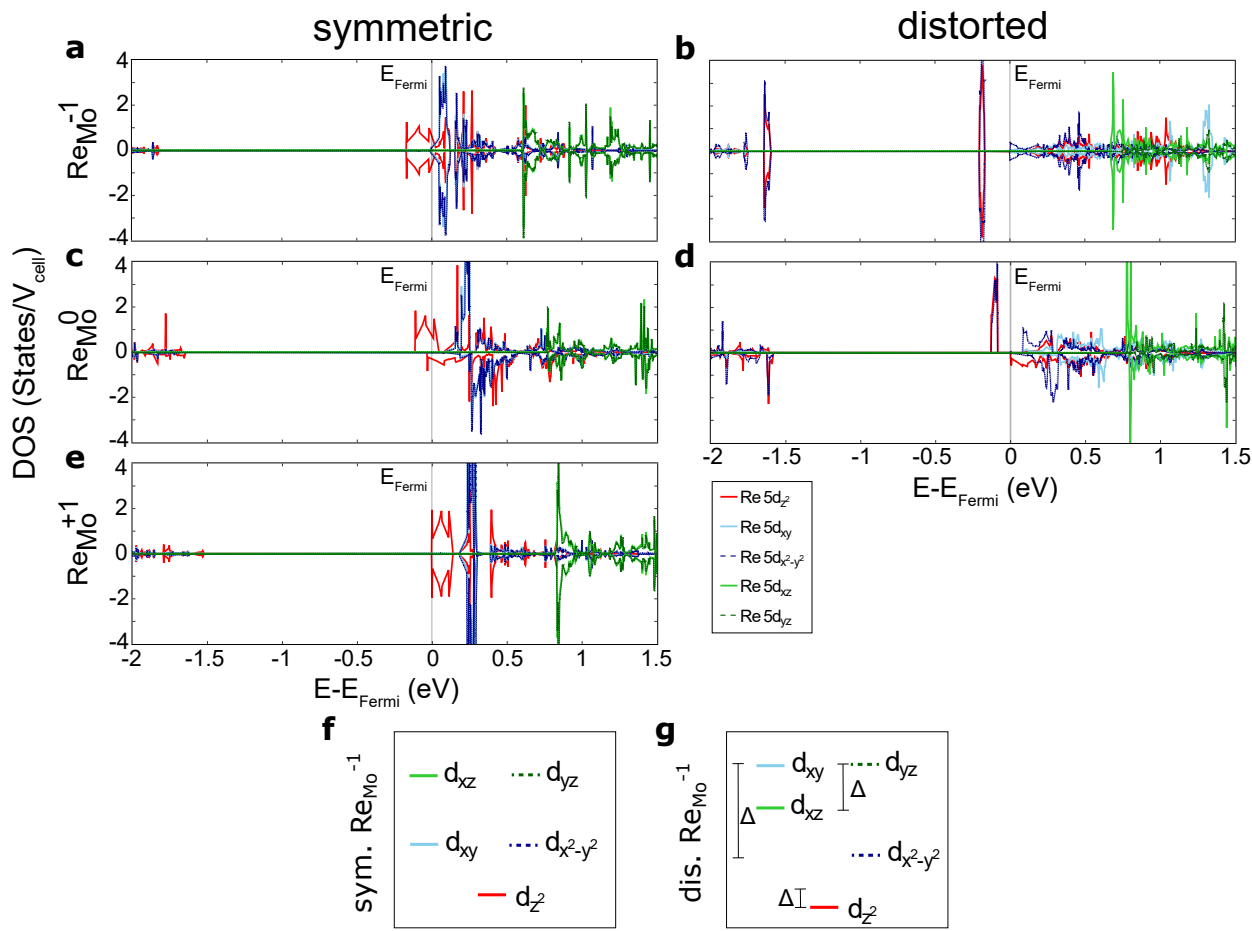}
\caption{\label{fig:Re_pDOS}
\textbf{Projected density of states of Re$_\text{Mo}$ in different charge states and geometries.} 
\textbf{a,b}, calculated pDOS of Re$_\text{Mo}^{-1}$ in symmetric  (a) and distorted (b) configuration.   
\textbf{c,d}, calculated pDOS of Re$_\text{Mo}^{0}$ in symmetric (c) and distorted (d) configuration. For Re$_\text{Mo}^{-1}$ and Re$_\text{Mo}^{0}$ the distorted configuration is the ground state.
\textbf{e}, calculated pDOS of Re$_\text{Mo}^{+1}$ in symmetric configuration.
\textbf{f,g}, energy level diagrams of Re$_\text{Mo}^{-1}$ $d$ orbitals in symmetric (f) and distorted (g) configuration. 
}
\end{figure*}

The DFT calculated pDOS of Re$_\text{Mo}$ in different charge states and geometries are shown in \figref{fig:Re_pDOS}. For both the neutral and negative charge state a strong reconfiguration of the electronic structure of the frontier defect orbitals are observed upon distortion. The distortion leads to a net energy gain of the total energy of the system. The initial symmetric geometry shows an orbital degeneracy between $d_{xy}$, $d_{x^2-y^2}$ and $d_{xz}$, $d_{yz}$ in the unoccupied spectrum. The occupied defect orbital in the Re$_\text{Mo}^{-1}$ and Re$_\text{Mo}^{0}$ states is well projected onto the Re 5$d_{z^2}$ orbital, which is not degenerate with any other states. Therefore, a regular Jahn-Teller effect cannot appear. However, we do see the $d_{z^2}$ lower in energy upon occupation. Therefore we propose a pseudo Jahn-Teller (pseudo-JTE) effect to be primarily responsible for the distortion of the defect. \\

In the case of the positively charged defect, there is no distorted structure observed because we do not find a stable distorted structure for this charge state. This is expected since none of the closely spaced defect states close to the conduction band are occupied. The Re impurity in the +1 state is isovalent to the Mo it replaces, hence both on a qualitative and quantitative point of view, no qualitative change in bonding behavior is expected in line with experimental observations.\\

\clearpage

\begin{figure*}[h!]
\includegraphics[width=\textwidth]{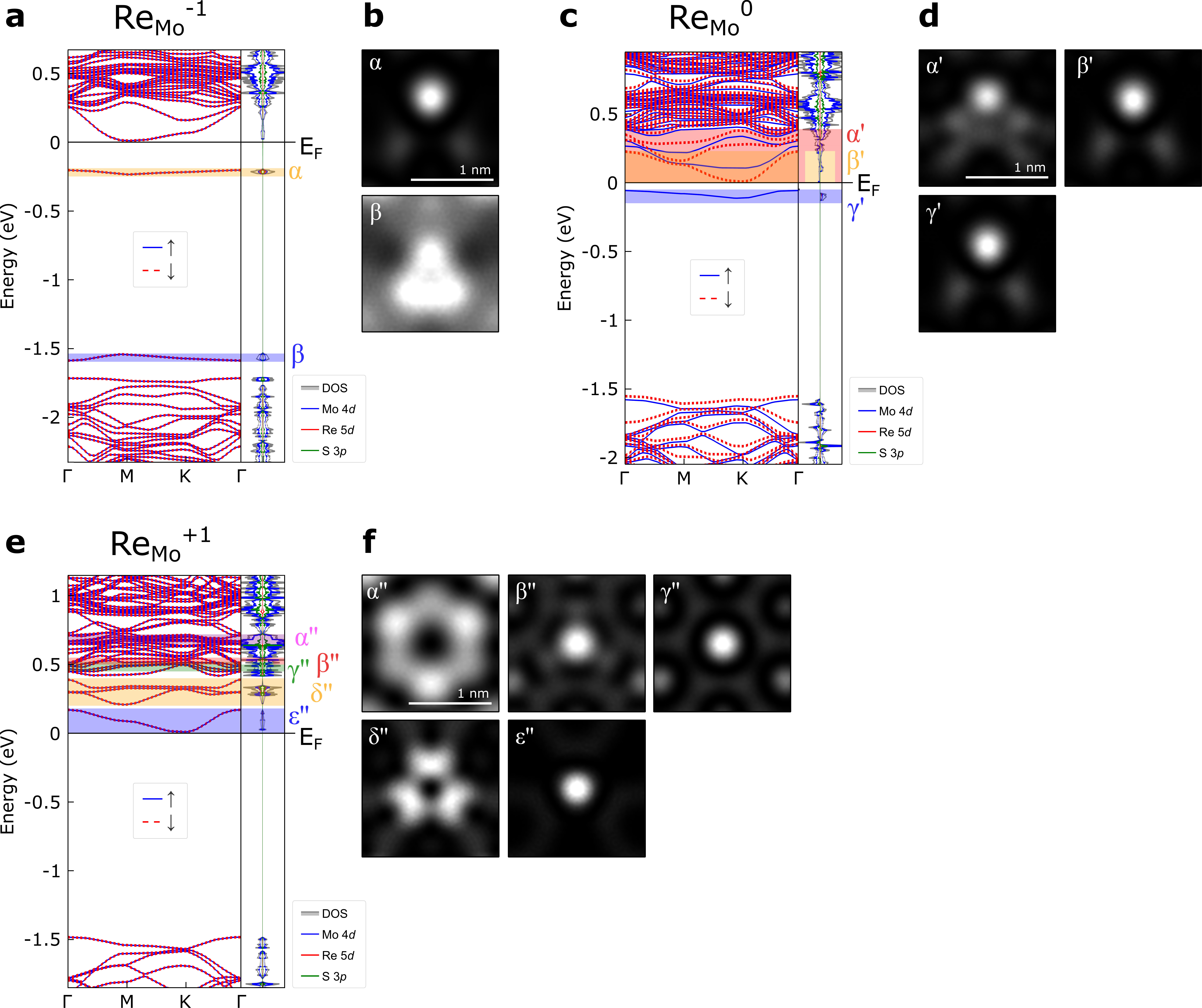}
\caption{\label{fig:Re_DOS_maps_all}
\textbf{DFT calculated band structures, projected density of states (pDOS) and simulated d$I$/d$V$ maps of Re$_\text{Mo}$ in different charge states.} 
\textbf{a,c,e}, band structure (left) and pDOS (right) of Re$_\text{Mo}^{-1}$ (a), Re$_\text{Mo}^{0}$ (c), and Re$_\text{Mo}^{+1}$ (e).
\textbf{b} simulated dI/dV maps of Re$_\text{Mo}^{-1}$ at the corresponding energy marked in (a).
\textbf{d} simulated dI/dV maps of Re$_\text{Mo}^{0}$ at the corresponding energy marked in (c).
\textbf{f} simulated dI/dV maps of Re$_\text{Mo}^{+1}$ at the corresponding energy marked in (e).
}
\end{figure*}

In \figref{fig:Re_DOS_maps_all} the calculated band structure and a selection of frontier Re$_\text{Mo}$ orbitals are shown. For Re$_\text{Mo}^{-1}$ we find a fully occupied orbital $\alpha$ exhibiting a clear two-fold symmetry (\figref{fig:Re_DOS_maps_all}b). An equivalent orbital shape is also found for the Re$_\text{Mo}^{0}$, labelled $\gamma'$. However, this orbital is only singly occupied leading to a spin-polarized band structure and a corresponding unoccupied orbital shape in the orthogonal spin channel ($\beta'$). This orbital is in excellent agreement with experimental d$I$/d$V$ maps shown in Fig.~5b. The positively charged Re$_\text{Mo}^{+1}$ exhibits only three-fold symmetric orbitals as expected, which are in good agreement with the experimental d$I$/d$V$ maps shown in \figref{fig:Re_Mo_dIdV_+1}c.\\

The prominent three-lobal orbital shape (cf. resonance E in Fig.~5b) could not be observed in the calculations. We suspect that this could in fact be due to a dynamic effect not considered in our calculations: at higher negative bias, inelastic tunnel electrons may drive the transition between the three equivalent distorted geometries that may create a dynamic superposition of several geometric configurations. Following this rationale, we would expect the probed orbital to mimic the superposition of the defect orbital $\gamma'$ with its 120$^\circ$ and 240$^\circ$ copies that may also include the fully symmetric geometry as a transition state in the APES. Qualitative comparison yielded indeed a similar shape to what has been observed experimentally. In-depth molecular dynamics simulations are needed to test this hypothesis and to understand the exact pathways of geometric transitions that may be described as a dynamic JTE. 

\clearpage

\bibliographystyle{nature_bsc}
\bibliography{references.bib}